\documentclass[acmsmall]{acmart}

\renewcommand\footnotetextcopyrightpermission[1]{} 

\AtBeginDocument{%
  }

\newcommand{\ourmodel}{\textsc{PaR}}
\newcommand{\ourdata}{\textsc{Defects4DS}}

\setcopyright{acmcopyright}
\copyrightyear{2018}
\acmYear{2018}
\acmDOI{XXXXXXX.XXXXXXX}

\usepackage{diagbox}
\usepackage{multirow}
\usepackage{graphicx}
\usepackage{booktabs}
\usepackage{subcaption}
\usepackage{pifont}
\usepackage{tcolorbox}
\usepackage{mdframed}
\usepackage{xcolor,colortbl}
\usepackage{tabularx}

\definecolor{c1}{cmyk}{0,0.6175,0.8848,0.1490}
\definecolor{c2}{cmyk}{0.1127,0.6690,0,0.4431}
\definecolor{c3}{cmyk}{0.3081,0,0.7209,0.3255}
\definecolor{c4}{cmyk}{0.6765,0.2017,0,0.0667}
\definecolor{c5}{cmyk}{0,0.8765,0.7099,0.3647}

\definecolor{lightgrey}{rgb}{0.93,0.93,0.93}

\newtcbox{\hlprimarytab}{on line, rounded corners, box align=base, colback=c3!10,colframe=white,size=fbox,arc=3pt, before upper=\strut, top=-2pt, bottom=-4pt, left=-2pt, right=-2pt, boxrule=0pt}
\newtcbox{\hlsecondarytab}{on line, box align=base, colback=red!10,colframe=white,size=fbox,arc=3pt, before upper=\strut, top=-2pt, bottom=-4pt, left=-2pt, right=-2pt, boxrule=0pt}
\newtcbox{\hlthirdtab}{on line, rounded corners, box align=base, colback=c4!10,colframe=white,size=fbox,arc=3pt, before upper=\strut, top=-2pt, bottom=-4pt, left=-2pt, right=-2pt, boxrule=0pt}

\newcommand{\dashifted}{\raisebox{0.5\depth}{\tiny$\downarrow$}}
\newcommand{\uashifted}{\raisebox{0.5\depth}{\tiny$\uparrow$}}
\newcommand{\da}[1]{{\scriptsize\hlsecondarytab{\dashifted{#1}}}}
\newcommand{\ua}[1]{{\scriptsize\hlprimarytab{\uashifted{#1}}}}
\newcommand{\percent}[1]{{\small\hlthirdtab{#1}}}

\begin{document}

\title{Peer-aided Repairer: Empowering Large Language Models to Repair Advanced Student Assignments}

\author{Qianhui Zhao}
\authornote{Both authors contributed equally to this research.}
\email{zhaoqianhui@buaa.edu.cn}
\author{Fang Liu}
\authornotemark[1]
\email{fangliu@buaa.edu.cn}
\affiliation{%
  \institution{Beihang University}
  \streetaddress{37 Xueyuan Road, Haidian District}
  \city{Beijing}
  \country{China}
  \postcode{100191}
}

\author{Li Zhang}
\affiliation{%
  \institution{Beihang University}
  \streetaddress{37 Xueyuan Road, Haidian District}
  \city{Beijing}
  \country{China}
  \postcode{100191}
}
\email{lily@buaa.edu.cn}

\author{Yang Liu}
\affiliation{%
  \institution{Beihang University}
  \streetaddress{37 Xueyuan Road, Haidian District}
  \city{Beijing}
  \country{China}
  \postcode{100191}
}
\email{liuyang26@buaa.edu.cn}

\author{Zhen Yan}
\affiliation{%
  \institution{Beihang University}
  \streetaddress{37 Xueyuan Road, Haidian District}
  \city{Beijing}
  \country{China}
  \postcode{100191}
}
\email{yanz@buaa.edu.cn}

\author{Zhenghao Chen}
\affiliation{%
  \institution{Beihang University}
  \streetaddress{37 Xueyuan Road, Haidian District}
  \city{Beijing}
  \country{China}
  \postcode{100191}
}
\email{czh20020503@buaa.edu.cn}

\author{Yufei Zhou}
\affiliation{%
  \institution{Beihang University}
  \streetaddress{37 Xueyuan Road, Haidian District}
  \city{Beijing}
  \country{China}
  \postcode{100191}
}
\email{zhouyufei@buaa.edu.cn}

\author{Jing Jiang}
\affiliation{%
  \institution{Beihang University}
  \streetaddress{37 Xueyuan Road, Haidian District}
  \city{Beijing}
  \country{China}
  \postcode{100191}
}
\email{jiangjing@buaa.edu.cn}

\author{Ge Li}
\affiliation{%
  \institution{Peking University}
  \streetaddress{No.5 Yiheyuan Road, Haidian District}
  \city{Beijing}
  \country{China}
  \postcode{100871}
}
\email{lige@pku.edu.cn}



\begin{abstract}

Automated generation of feedback on programming assignments holds significant benefits for programming education, especially when it comes to advanced assignments.
Automated Program Repair techniques, especially Large Language Model based approaches, have gained notable recognition for their potential to fix introductory assignments. However, the programs used for evaluation are relatively simple. It remains unclear how existing approaches perform in repairing programs from higher-level programming courses.
To address these limitations, we curate a new advanced student assignment dataset named \ourdata{} from a higher-level programming course. Subsequently, we identify the challenges related to fixing bugs in advanced assignments. Based on the analysis, we develop a framework called \ourmodel{} that is powered by the LLM. \ourmodel{} works in three phases: Peer Solution Selection, Multi-Source Prompt Generation, and Program Repair. Peer Solution Selection identifies the closely related peer programs based on lexical, semantic, and syntactic criteria. Then Multi-Source Prompt Generation adeptly combines multiple sources of information to create a comprehensive and informative prompt for the last Program Repair stage. The evaluation on \ourdata{} and another well-investigated ITSP dataset reveals that \ourmodel{} achieves a new state-of-the-art performance, demonstrating impressive improvements of 19.94\% and 15.2\% in repair rate compared to prior state-of-the-art LLM- and symbolic-based approaches, respectively.
\end{abstract}

\maketitle

\section{Introduction}

Programming education occupies an important position in developing talents in the software engineering field, which has received wide attention in the past decades \cite{gulwani2018clara,Ahmed2021VerifixVR}. Providing feedback on programming assignments for students is an essential part and also a tedious task, which requires substantial effort by the teachers and the teaching assistants. With the development of computer science and digital society, the demand for programming education increased sharply, spawning Massive Open Online Courses (MOOCs) that teach programming \cite{gulwani2018clara,Masters2011ABG}. Under such circumstances, providing customized and timely feedback to a large number of students becomes more pressing, even impossible with the massive number of students. 

There has been increased interest in applying Automated Program Repair (APR) techniques for providing feedback on student programming assignments. APR tools can automatically generate patches to correct code errors by reasoning about the code semantics based on the given specification \cite{Gazzola2018AutomaticSR,Goues2019AutomatedPR}. A variety of approaches have been proposed across the years, including search-based methods \cite{le2011genprog,liu2019tbar}, semantic-based tools \cite{nguyen2013semfix,mechtaev2016angelix,le2017jfix}, and Deep Learning (DL)-based techniques \cite{gupta2017deepfix,jiang2021cure}. 
Based on existing advanced APR techniques, various approaches have been introduced for correcting students' mistakes in their programming assignments \cite{gulwani2018clara,Ahmed2021VerifixVR,singh2013AutoGrader,rolim2017REFAZER}, aiming at facilitating the learning procedure \cite{hu2019re}.
The search-based and semantic-based methods \cite{singh2013AutoGrader,gulwani2018clara,wang2018search,Ahmed2021VerifixVR} demand extensive engineering efforts, including program analysis or repair experience and custom repair strategies tailored to the language domain in which students complete their assignments. 
Moreover, these methods have limited generality and are specifically tailored for simple syntax components, excluding more intricate elements such as pointers, structures, and multidimensional arrays. 
DL-based approaches alleviate some engineering challenges by capturing various bug-fixing patterns from large code corpora. However, these approaches generally rely on large quantities of training data \cite{gupta2017deepfix,pu2016sk_p,yasunaga2021break}, and are still limited to generating small patches due to the complexity of semantic reasoning, and search space explosion.

Recent advancements in the development of Large Language Models (LLMs) provide an alternative solution for bug repair that does not necessitate experts with program analysis/repair experience. Most recently, researchers have started to directly leverage advanced LLMs for APR \cite{joshi2023repair,fan2023automated,jiang2023impact,xia2022practical}, as well as repairing the student programming assignments \cite{zhang2022repairing,joshi2023repair}. These models have shown promising results in student assignment bug fixing, and even outperform previous state-of-the-art APR techniques. However, these approaches suffer from the following limitations. 
The first limitation is the use of relatively simple programs for evaluation. The datasets they utilized are typically comprised of assignments from introductory programming courses, which tend to be shorter and simpler, averaging around 20 lines of code and without complex grammatical components. Consequently, it remains uncertain how these evaluated approaches would act when applied to more advanced programming courses, which contain assignments with more intricate grammatical components and logic.
Furthermore, many of the approaches utilize the peer solution to assist the repair process \cite{joshi2023repair,zhang2022repairing}.
However, the selection process primarily relies on the execution of test suites or compiler messages, overlooking the implementation details and underlying code structure. Therefore, there may be discrepancies between the chosen peer solution and the students' faulty code, making them less effective in fixing the bugs present in their own code.
The limitations in both evaluation and example selection undermine the effectiveness and robustness of the approaches, hindering their ability to provide comprehensive and reliable solutions for complex programming issues.

To address the above issues, \ding{182} we first curate a new student assignment dataset named \ourdata{} based on a higher-level programming course, including programs with increased complexity, longer lengths, and a variety of structures. Our goal is to obtain valuable insights into the performance and limitations of the evaluated APR approaches when applied to programs from advanced programming courses. These programs feature complex grammatical components and logic, making the bugs within them more challenging to locate and fix.
\ding{183} Then, we outline the challenges involved in repairing bugs in advanced student assignments, drawing upon our analysis of the characteristics of \ourdata{} and comparing them to an introductory programming assignment dataset. 
\ding{184} To address the challenges, we propose Peer-aided Repairer (\ourmodel{}), which is a novel framework for advanced student assignment repair powered by a large language model. \ourmodel{} works in three phases: \textit{Peer Solution Selection}, \textit{Multi-Source Prompt Generation}, and \textit{Program Repair}. 
During the \textit{Peer Solution Selection} stage, \ourmodel{} identifies the closely related peer programs based on a mix of lexical, semantic, and syntactic criteria. Then during the \textit{Multi-Source Prompt Generation} stage, \ourmodel{} combines multiple sources of information to create a comprehensive and informative prompt $\mathcal{P}$, including the peer solution obtained in the previous step, program description, IO-related information, buggy code, \textit{etc}. Finally, during the \textit{Program Repair}
stage, the prompt generated by the previous stage is fed to the LLM, which then produces fixed code that is subsequently evaluated for correctness. 

We evaluate \ourmodel{} on students' assignment bug fixing task using both our proposed \ourdata{} dataset and another widely used introductory student assignment benchmark dataset ITSP \cite{yi2017itsp}. The evaluation results demonstrate that \ourmodel{} achieves superior performance in repairing both advanced and introductory student assignments, obtaining 301 correct fixes on \ourdata{} and 251 correct fixes on the ITSP dataset, showcasing a remarkable improvement of 19.94\% and 15.2\% in repair rate compared to LLM-based model \cite{li2023starcoder} and the powerful symbolic approach Verifix \cite{Ahmed2021VerifixVR}.
Further analysis also confirms the effectiveness of each component in \ourmodel{}.
The contributions of this paper can be summarized as follows:
\begin{itemize}
    \item  We propose \ourdata{}, a new dataset that contains 682 submissions from 4 programming assignments of a higher-level programming course. Additionally, we meticulously labeled every buggy code in the \ourdata{} dataset, offering comprehensive bug information across four dimensions. Moreover, we also labeled the buggy code in the introductory programming assignment dataset (ITSP), enabling an in-depth comparison between \ourdata{} and ITSP.
    \item We design a novel repair framework \ourmodel{}, which empowers large language models to repair student assignments with an innovative peer solution selection strategy and a multi-source prompt generation method.
    \item We conduct a comprehensive evaluation of \ourmodel{} in comparison to both large language models and a symbolic APR tool across \ourdata{} and ITSP datasets. The evaluation results demonstrate that \ourmodel{} consistently outperforms all the baseline approaches by a large margin.
    \item We open-source \ourdata{} along with the corresponding bug information, and the replication package of \ourmodel{} at \url{https://figshare.com/s/0f7481a285fcab11c2f9}. 
\end{itemize}

\section{Related Works\&Background}

\subsection{Large Language Models}
Large language models (LLMs) are a category of large-scale models that have been pre-trained on a massive textual corpus with self-supervised pre-training objectives \cite{devlin2018bert, nijkamp2022codegen}. 
Most LLMs are built on the Transformer architecture \cite{vaswani2017attention}, and can be categorized into three groups: encoder-only models \cite{devlin2018bert}, decoder-only models \cite{gpt3}, and encoder-decoder models \cite{Raffel2020t5}.
Later, researchers further incorporated reinforcement learning to align LLMs with human intent. One notable LLM, OpenAI's ChatGPT \cite{chatgpt}, applies reinforcement learning from human feedback (RLHF) to optimize the model. This approach has garnered significant attention due to its remarkable ability to tackle a wide range of tasks. 
In the meanwhile, various open-source LLMs have emerged in the AI community. Notable contributions include GPT-NeoX \cite{black2022gpt}, GPT-J \cite{wang2021gpt-j}, InCoder \cite{fried2022incoder}, CodeGen \cite{nijkamp2022codegen}, LLaMA \cite{touvron2023llama}, StarCoder \cite{li2023starcoder}, Code Llama \cite{roziere2023codellama}, \textit{etc}. These models also have showcased remarkable progress in various tasks. 
It was worth noting that LLMs are highly effective in few-shot learning scenarios, which can successfully perform tasks for which they were not explicitly trained by providing only a few (or no) examples during the inference. This approach is commonly known as prompt-based learning \cite{liu2023pre}. A prompt generally refers to a predefined textual template that is given as an input instruction to the LLM, and it typically consists of a query and may include a few examples (also known as shots) of the task. LLM can generate results of corresponding tasks based on the prompt.
There have been recent studies exploring ChatGPT and other open-source LLMs' potential in software engineering tasks, such as program repair \cite{fan2023automated,xia2023keep}, code generation \cite{dong2023self}, unit test generation \cite{yuan2023no}, \textit{etc}. 

\subsection{Automated Program Repair}
\subsubsection{General Purpose Program Repair}
Automated Program Repair (APR) techniques are capable of identifying and fixing software bugs or defects without human intervention, aiming to reduce the effort required to improve software quality. A variety of APR approaches have been proposed across the years, including search-based methods, semantic-based approaches, and Deep Learning-based techniques. The search-based approaches first produce patches with predefined code transformation operators, and then search for correct patches among the pre-defined space \cite{le2011genprog,liu2019tbar}. The search-based repair approaches demonstrate the ability to handle large programs. However, they may struggle when confronted with expansive search spaces.
The semantic-based tools usually correct code errors via symbolic execution \cite{nguyen2013semfix,mechtaev2016angelix,le2017jfix}. Specifically, they need to build a constraint that a program must meet to pass a given test suite, and then solve the constraint to generate the patches. 
Deep Learning-based APR approaches predominantly rely on Neural Machine Translation (NMT) techniques \cite{gupta2017deepfix,jiang2021cure,zhu2021recoder}, which consider the program repair as a translation task, where the goal is to transform defective code into correct code.
However, the effectiveness of NMT-based APR tools heavily relies on the quality and abundance of training data. 
More recently, researchers have begun to directly harness the capabilities of LLMs for APR \cite{xia2023keep,xia2022practical,fan2023automated},
which have surpassed the existing state-of-the-art methods across various program repair tasks. For instance, ChatRepair \cite{xia2023keep}, which is an APR tool powered by ChatGPT, fixes 114 bugs on Defects4J 1.2, with 15 more than the previous state-of-the-art APR tools. These tools work by first creating an input prompt with the original buggy code as well as the task description. Then they query the LLM to either fill in the correct code at the bug location or generate a new code snippet as the patch.

\subsubsection{Programming Assignment Program Repair}
Researchers have extended the application of APR to student programs in education, 
providing feedback to help students correct their programs \cite{singh2013AutoGrader,gulwani2018clara,wang2018search}. 
AutoGrader \cite{singh2013AutoGrader} is an early feedback generation system that takes an erroneous student program, a reference solution, and a set of potential corrections provided by the instructor as inputs. It employs program synthesis techniques to find minimal corrections. 
CLARA \cite{gulwani2018clara} and sk\_p \cite{pu2016sk_p} employed clustering and machine learning techniques to generate repairs for the assignment data. However, these repairs frequently exhibit imprecision and lack minimality, thereby diminishing the overall quality of the feedback.
Verifix \cite{Ahmed2021VerifixVR} aims to generate verifiable correct program repairs by aligning their assignments with a reference solution in terms of control flow. They utilize failed verification attempts to obtain potential repairs through the MaxSMT solver.
More recently, researchers have begun to employ LLMs to fix bugs in student assignments. \citet{zhang2022repairing} built an APR system based on Codex \cite{chen2021evaluatingCodex} for introductory Python programming assignments. 
The empirical results on 286 Python programs demonstrated that their system outperforms other tools. \citet{xia2023keep} proposed a conversation-driven APR approach based on ChatGPT \cite{chatgpt}, which utilized instant feedback to perform patch generation in a conversational style. They evaluated both their approach on the Defects4J \cite{just2014defects4j} and QuixBugs \cite{lin2017quixbugs} dataset, and the results demonstrated their approach obtains the superior result on both datasets.
However, there is still limited understanding regarding LLM's effectiveness in fixing bugs in advanced student programming assignments.

\subsubsection{Benchmarks for Programming Assignment Program Repair}
Existing programming assignment-oriented approaches to APR focus on fixing bugs in C and Python.
The widely used datasets in C are IntroClass \cite{le2015manybugsandintroclass} and ITSP \cite{yi2017itsp}. IntroClass is collected from an introductory C programming class with an enrollment of about 200 students. It consists of 998 submitted programs with 572 unique defects out of 1143 defects in total (sometimes the students submitted identical code, thus there may exist identical defects). The average line of code is relatively short with around 20 lines of code. ITSP consists of 661 buggy-repaired program pairs submitted by students from an introductory C programming course for 74 assignments.
For Python, the dataset used in \citet{yang2019refactory} is collected from an introductory Python programming course credited by 361 students, and contains 2442 correct and 1783 buggy program attempts with an average lines of code of 14. Another dataset used in \citet{gulwani2018clara} consisting of student attempts from MITx MOOC is not publicly available.

\section{Dataset}
To facilitate the evaluation of higher-level programming assignment repair models, we collect a new dataset called \ourdata{} from the Data Structures course in a college. 
Subsequently, we gather comprehensive bug information from 4 dimensions to facilitate further analysis of the buggy code. Furthermore, we conduct a meticulous examination of the characteristics of the buggy code within our dataset and the ITSP \cite{yi2017itsp} dataset, and compare their respective features. This analysis helps us outline the specific challenges involved in repairing the bugs in advanced student programming assignments. The subsequent content will present each of these processes in detail.

\subsection{Data Collection}
We collected the dataset from a Data Structures course in a college, which is a foundational professional course for computer science students. Compared to the introductory programming courses, the Data Structures course emphasizes problem-solving, algorithmic thinking, and optimization techniques, preparing students to work with large and complex code files. To finish the assignments, students will explore various data structures (such as linked lists, stacks, queues, trees, \textit{etc.}).
In this course, students are tasked with writing C programs that meet the provided problem specifications, including detailed descriptions, input and output formats, test examples, and occasionally illustrations.
For each assignment, students have the flexibility to make an unlimited number of submissions until the specified deadline or until their submission successfully passes all the provided test cases.

We gathered students' submissions from 4 assignments of different complexity levels. Each submission includes two code variations: a correct code that successfully passes all five test cases, and a buggy code that represents the student's last incorrect attempt.
Considering that most syntax errors can be detected by compilers and the inherent difficulty that students face in addressing semantic errors, we have exclusively retained buggy codes that contain semantic errors. Then we further narrow down our selection to entries where the discrepancy between the correct code and the buggy code is limited to a maximum of five lines. This ensures that the code can be repaired with a few edits, rather than necessitating a near-complete rewrite. Finally, the selection process has resulted in 682 submissions.

\subsection{Data Statistics \& Comparison with ITSP} \label{sec:data_statistics}

As shown in Table \ref{tab:programming problems},
the average and median number of lines of code in \ourdata{} is 55 and 78, respectively. These values are much higher compared to the datasets utilized in introductory programming courses \cite{yi2017itsp}, which exhibit average values of 22 and a median of 20. As a result, this disparity implies a higher level of complexity and difficulty in repairing the bugs present in \ourdata{}. 

Besides, we further analyze the distribution of grammatical components, including complex grammatical components and custom functions.
For complex grammatical components, we take the struct, pointer, and multi-dimensional array into consideration. These components are commonly utilized by students when completing programming assignments, as they play a crucial role in managing and organizing complex data structures.
According to the analysis results, shown in Table \ref{tab:programming problems}, these components occur in 263/682 (38.6\%) Defects4DS programs (the total count is less than the sum of three components due to the occurrence of multiple complex grammatical components in one piece of code), while never in ITSP used in Verifix. Additionally, the custom functions occur in 291/682 (42.7\%) Defects4DS programs, while only 20.5\% (70/341) in ITSP, implying the gap of fix difficulty. 

\subsection{Data Labeling}\label{data_labeling}

To streamline further analysis, we gather comprehensive bug information from four dimensions within the buggy code in both \ourdata{} and ITSP datasets: \textit{Bug Line}, \textit{Bug Type}, \textit{Repair Type}, and \textit{Bug Correlation}. During the labeling process, we designate the buggy code as the annotation object, utilizing the correct code as a reference. It's important to note that certain invalid changes are disregarded, such as deleting a defined but unused variable, replacing variable names, and making equivalent modifications (\textit{e.g.}, \verb|i+=1| and \verb|i=i+1|), and so forth. 
The labeling and analysis were carried out by four of the authors, all of whom possess over three years of experience in C programming.
The corresponding detailed information for each dimension is outlined below. 

\begin{table}[t]
    \centering\scriptsize
    \setlength{\abovecaptionskip}{0.1cm}
    \caption{Statistics of \ourdata{} and ITSP. For ITSP, lab 3-6 are collected following Verifix \cite{Ahmed2021VerifixVR}, where the statistics are calculated based on their GitHub repositories. CGC stands for complex grammatical components, CF stands for custom function, and M-Array stands for multi-dimensional array.}
    \addtolength{\tabcolsep}{-2pt}
    \begin{tabular}{llp{3cm}cccccccc}
    \toprule
    \multirow{2}{*}{\textbf{Dataset}} & \multirow{2}{*}{\textbf{ID}} & \multirow{2}{*}{\textbf{Task Description}} & \multirow{2}{*}{\textbf{\#Programs}} & \multicolumn{3}{c}{\textbf{\#Lines of Code}} & \multicolumn{3}{c}{\textbf{\#CGC \textit{Prop}.}} & \multirow{2}{*}{\textbf{\#CF \textit{Prop}.}} \\
    \cmidrule(lr){5-7} \cmidrule(lr){8-10}
     & & & & \textbf{Avg.} & \textbf{Median} & \textbf{Max} & \textbf{Struct} & \textbf{Pointer} & \textbf{M-Array} \\
    \midrule
    \multirow{5}{*}{\ourdata{}} & Prob.1 & generation of full permutations & 123 & 40 & 35 & 85 & - & 21 & 1 & 119\\
     & Prob.2 & expression calculation & 108 & 68 & 63 & 245 & 1 & 13 & - & 34\\
     & Prob.3 & conversion between decimal form and scientific notation & 197 & 54 & 52 & 123 & - & 12 & - & 16\\
     & Prob.4 & continuous line segment & 254 & 58 & 54 & 150 & 193 & 92 & 29 & 122\\
    \cmidrule{2-11}
     & Overall & - & 682 & 55 & 78 & 245 & 194 & 138 & 30 & 291\\
    \midrule
    \multirow{5}{*}{ITSP} & Lab-3 & simple expressions, printf, scanf & 63 & 13 & 13 & 22 & - & - & - & -\\
     & Lab-4 & conditionals & 117 & 21 & 19 & 83 & - & - & - & -\\
     & Lab-5 & loops, nested loops & 82 & 23 & 24 & 57 & - & - & - & 45\\
     & Lab-6 & integer arrays & 79 & 28 & 28 & 57 & - & - & - & 25\\
     \cmidrule{2-11}
     & Overall & - & 341 & 22 & 20 & 83 & - & - & - & 70\\
    \bottomrule
    \end{tabular}
    \label{tab:programming problems}
    \vspace{-0.3cm}
\end{table}

\noindent \textbf{Bug Line.}
This dimension captures the line number(s) where the bug is present. In cases where the bug occurs within a code block, we prioritize labeling by the entire block range. For bugs involving the repair type of statement addition, we denote the bug line as a range within which new statements can be included.

\begin{figure*}[t]
    \centering
    \setlength{\abovecaptionskip}{0.1cm}
    \includegraphics[width=1.0\linewidth]{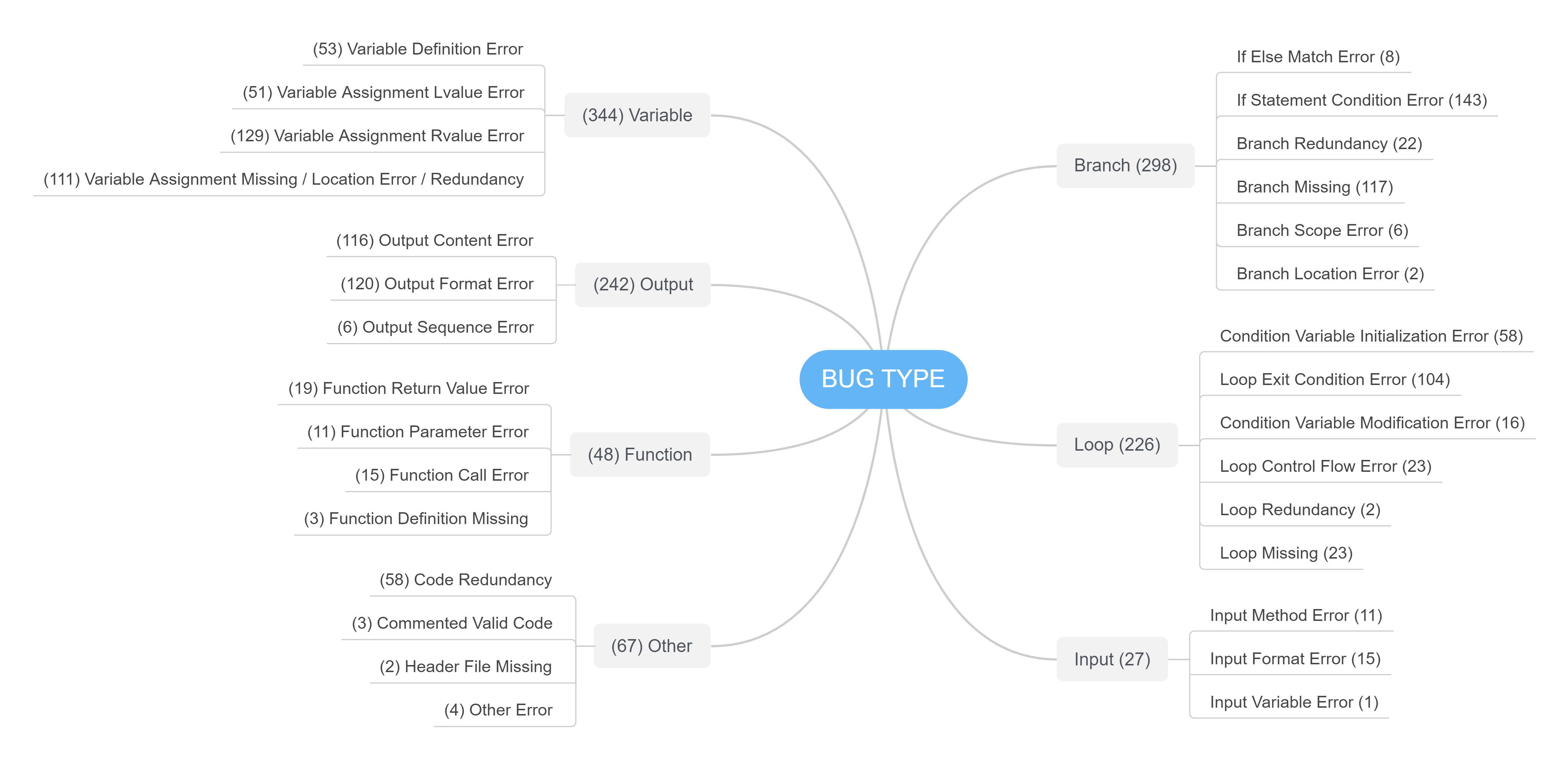}
    \caption{Bug type distribution of \ourdata{}. A single bug sometimes corresponds to multiple bug types.}
    \label{fig:bug types}
    \vspace{-0.3cm}
\end{figure*}

\noindent \textbf{Bug Type.}
This dimension categorizes the type of bug encountered. Our dataset divides bugs into 7 main categories and 30 subcategories, as shown in Figure \ref{fig:bug types}. The numbers in parentheses indicate the frequency of occurrence for each bug type in our dataset. As a single bug sometimes corresponds to multiple bug types, the total number of bug type occurrences is slightly more than the number of bugs. To establish this classification criteria, we conducted 4 rounds of labeling. 
In the \textit{first round}, we randomly selected a sample of 250 data points for analysis and labeling, achieving a 95\% confidence level and a 5\% confidence interval. The initial bug types were obtained by summarizing the labels from this sample data, merging bug types with similar meanings, and refining categories. 
In the \textit{second round}, the sample data was relabeled, and any additional bug types that were not covered in the initial categorization were added to form an updated version of bug types. 
The \textit{third round} involved relabeling the sample data and labeling the remaining data using the second version of bug types. Finally, refinement was conducted in the \textit{last round} to create the final version of bug types, and all the data was labeled accordingly. The detailed information of code book can be found in our replication package.

\noindent \textbf{Repair Type.}
This dimension categorizes the potential methods to fix a bug, which include:

\noindent \textit{(1) Statement Addition}: The fix involves adding new statements to the code.

\noindent \textit{(2) Statement Deletion}: The fix involves removing existing statements from the code.

\noindent \textit{(3) Statement Modification}: The fix involves making changes to the content of existing statements.

\noindent \textit{(4) Position Modification}: The fix involves changing the position or order of existing statements in the code.

\noindent These methods can also serve as indicators of the level of difficulty that APR models may encounter when attempting to fix a bug.

\noindent \textbf{Bug Correlation.}
Bug correlation indicates whether a bug is related to other bugs in the same file. 
We consider the related bugs that involve using the same variable or having impacts on the value of the same variable or expression. 
By recognizing bug correlation, APR tools might optimize their bug-fixing process since the fix for one bug may provide clues on fixing other related bugs.

\subsection{Data Analysis\&Challenges}
After the dataset is prepared, we analyze the characteristics of \ourdata{}.
Next, we compare these characteristics with those of ITSP and conclude by outlining the challenges involved in fixing bugs in advanced student assignments.

\noindent \textbf{Distribution of the number of bugs.}
We calculate the number of bugs in each piece of buggy code, and the results are shown in Table \ref{tab:number of bugs}. 
Overall, 58.4\% of the codes in \ourdata{} contain only one bug, about a quarter contain two bugs, and no more than a fifth contain three or more bugs. 
\ourdata{} contains 1181 bugs in total, with an average of 1.73 bugs per program.
Specifically, \textit{Prob.} 2 contains an average of 2.1 bugs per code, which is the most among all the problems.

\begin{table}[b]
    \centering\small
    \setlength{\abovecaptionskip}{0.1cm}
    \begin{tabular}{ccccccccc}
    \toprule
    \multirow{2}{*}{\textbf{Dataset}} & \multirow{2}{*}{\textbf{ID}} & \multicolumn{6}{c}{\textbf{\# Bugs per Code}} & \multirow{2}{*}{\textbf{Related Bugs}}\\
    \cline{3-8}
    & \multirow{2}{*}{} & 1 & 2 & 3 & 4 & 5 & >5\\
    \midrule
    \multirow{5}{*}{\ourdata{}}& Prob.1 & 72.4\% & 17.1\% & 5.7\% & 2.4\% & 1.6\% & 0.9\% & 53.0\%\\
    & Prob.2 & 42.6\% & 32.4\% & 5.6\% & 14.8\% & 4.6\% & - & 66.7\%\\
    & Prob.3 & 66.5\% & 23.9\% & 5.1\% & 2.5\% & 1.0\% & 1.0\% & 56.1\%\\
    & Prob.4 & 52.0\% & 26.0\% & 11.4\% & 5.5\% & 2.8\% & 2.4\% & 57.4\%\\
    \cline{2-9}
    & overall & 58.4\% & 24.8\% & 7.6\% & 5.6\% & 2.3\% & 1.3\% & 58.6\%\\
    \midrule
    \multirow{5}{*}{ITSP}& Lab-3 & 63.5\% & 14.3\% & 19.0\% & - & 3.2\% & - & 14.1\%\\
    & Lab-4 & 66.4\% & 15.0\% & 15.9\% & 0.9\% & 1.8\% & - & 8.8\%\\
    & Lab-5 & 68.1\% & 18.1\% & 11.1\% & 2.8\% & - & - & 15.5\%\\
    & Lab-6 & 63.0\% & 20.1\% & 9.6\% & 2.7\% & 2.7\% & 1.4\% & 14.7\%\\
    \cline{2-9}
    & overall & 65.4\% & 16.8\% & 14.0\% & 1.6\% & 1.9\% & 0.3\% & 12.7\%\\
    \bottomrule
    \end{tabular}
    \caption{Proportion of different numbers of bugs and related bugs for \ourdata{} and ITSP.}
    \label{tab:number of bugs}
    \vspace{-0.3cm}
\end{table}

\noindent \textbf{Proportion of related bugs.}
We use the calculation of the number of codes containing related bugs divided by the number of codes containing two or more bugs as the proportion of codes containing related bugs, since bug correlation is not involved for codes containing only one bug. Results indicate that all the problems in \ourdata{} have a large proportion of related bugs with an overall rate of 58.6\%, and \textit{Prob.} 2 has the highest value at 66.7\%.

\begin{figure}[t]
    \setlength{\abovecaptionskip}{0.1cm}
     \centering
     \begin{subfigure}[b]{0.45\textwidth}
        \setlength{\abovecaptionskip}{0cm}
         \centering
         \includegraphics[width=0.45\textwidth]{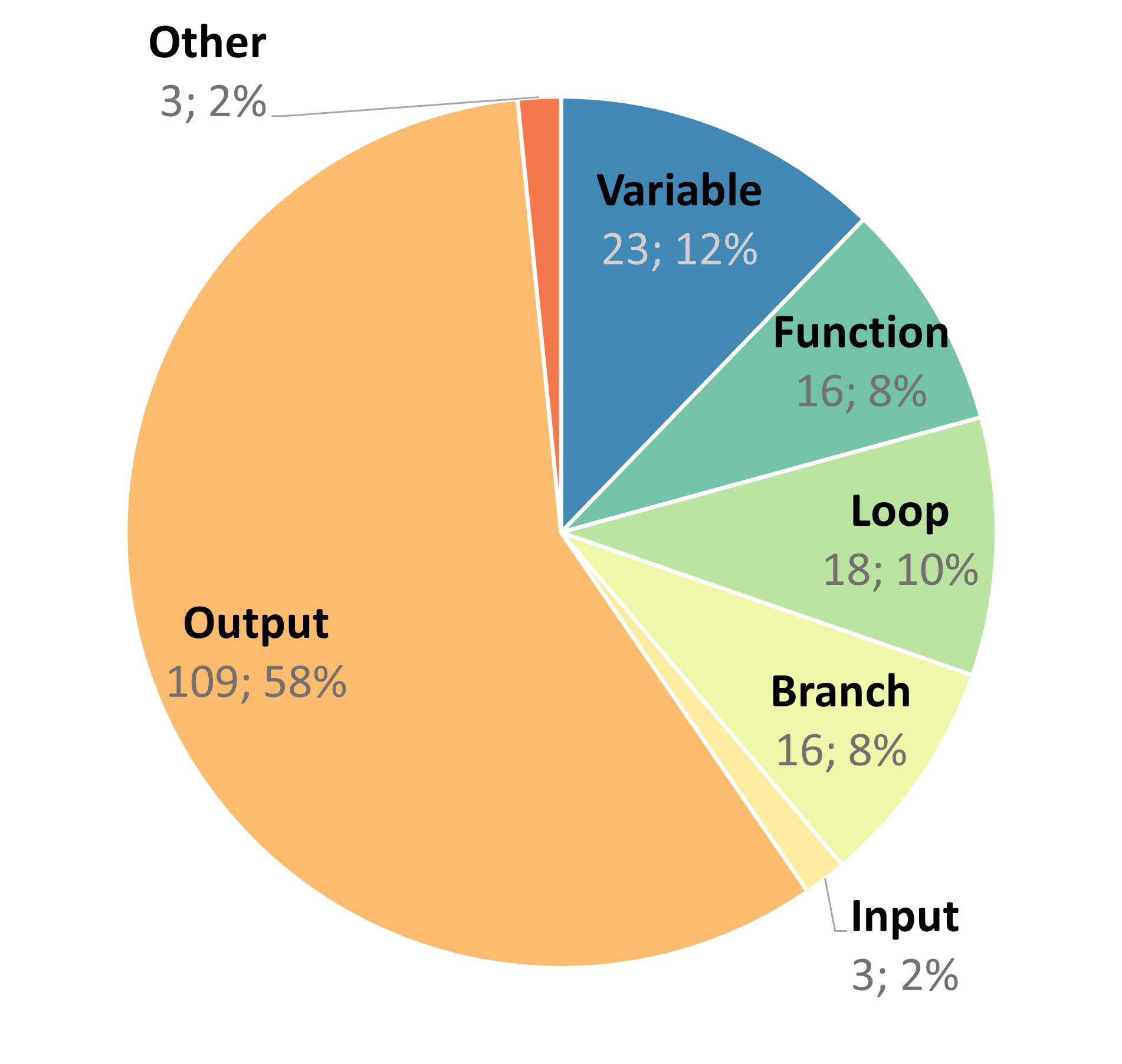}
         \includegraphics[width=0.45\textwidth]{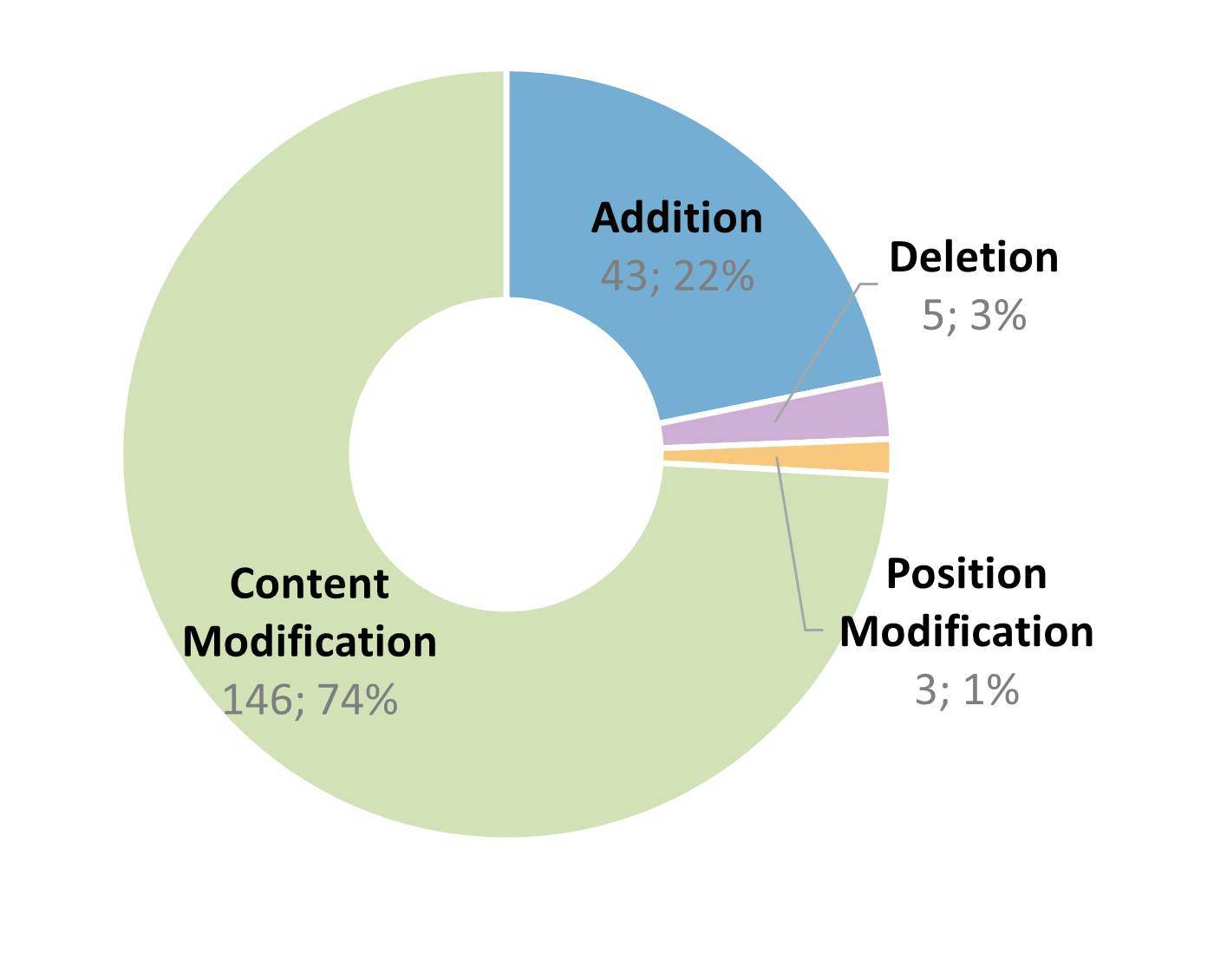}
         \caption{\textit{Prob.} 1}
         \label{fig:problem1}
     \end{subfigure}
     \begin{subfigure}[b]{0.45\textwidth}
        \setlength{\abovecaptionskip}{0cm}
         \centering
         \includegraphics[width=0.45\textwidth]{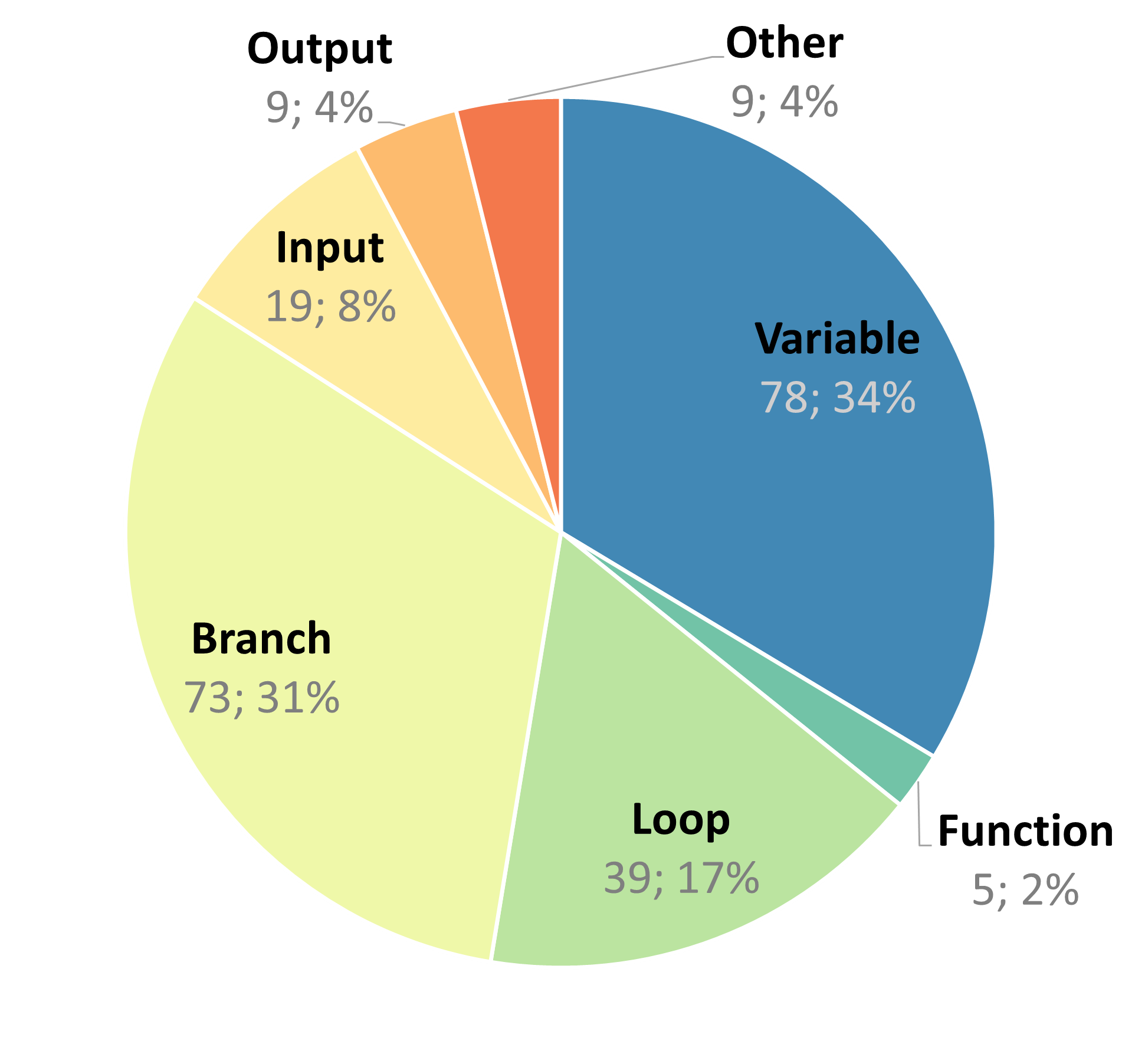}
         \includegraphics[width=0.45\textwidth]{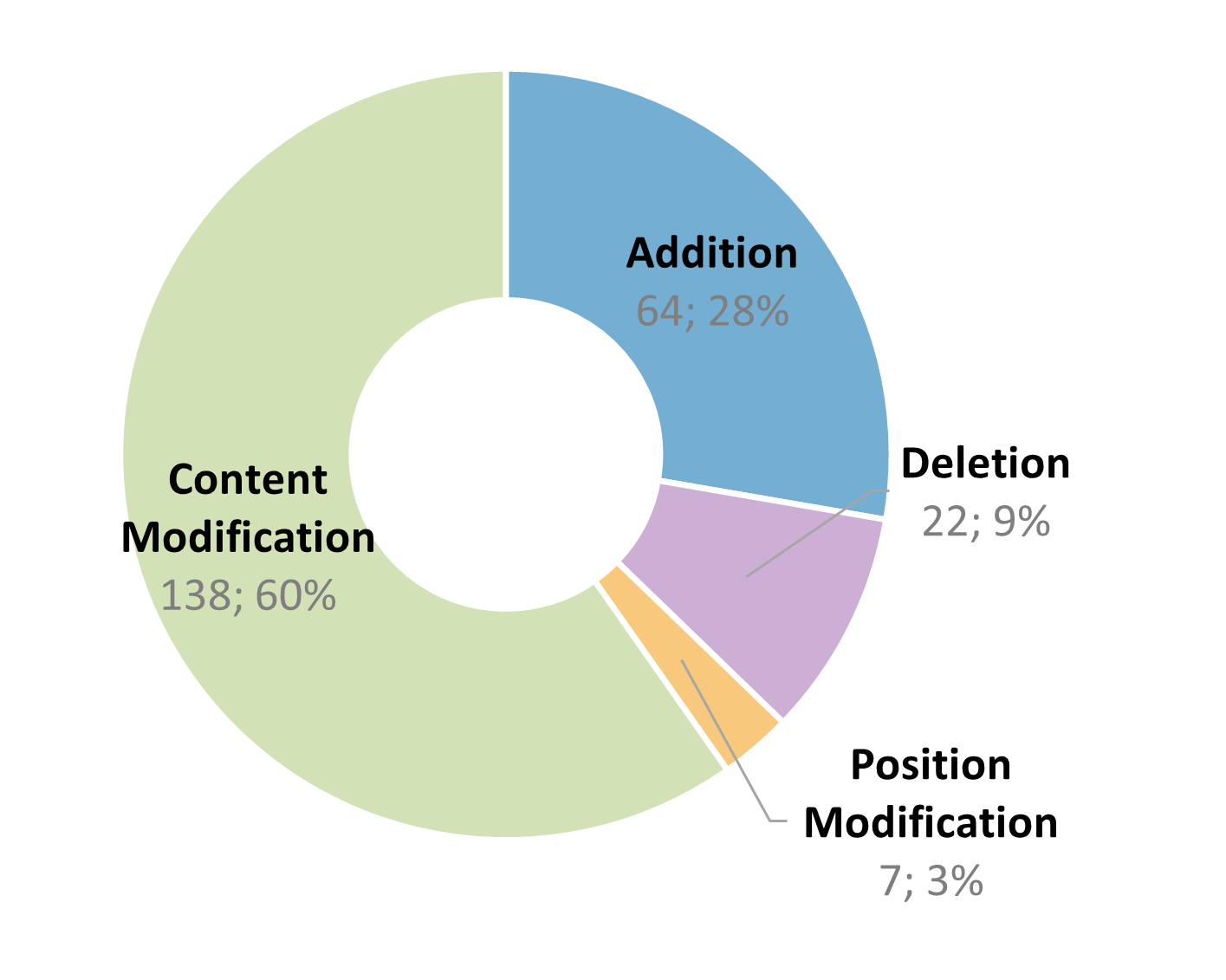}
         \caption{\textit{Prob.} 2}
         \label{fig:problem2}
     \end{subfigure}
     \begin{subfigure}[b]{0.45\textwidth}
        \setlength{\abovecaptionskip}{0cm}
         \centering
         \includegraphics[width=0.45\textwidth]{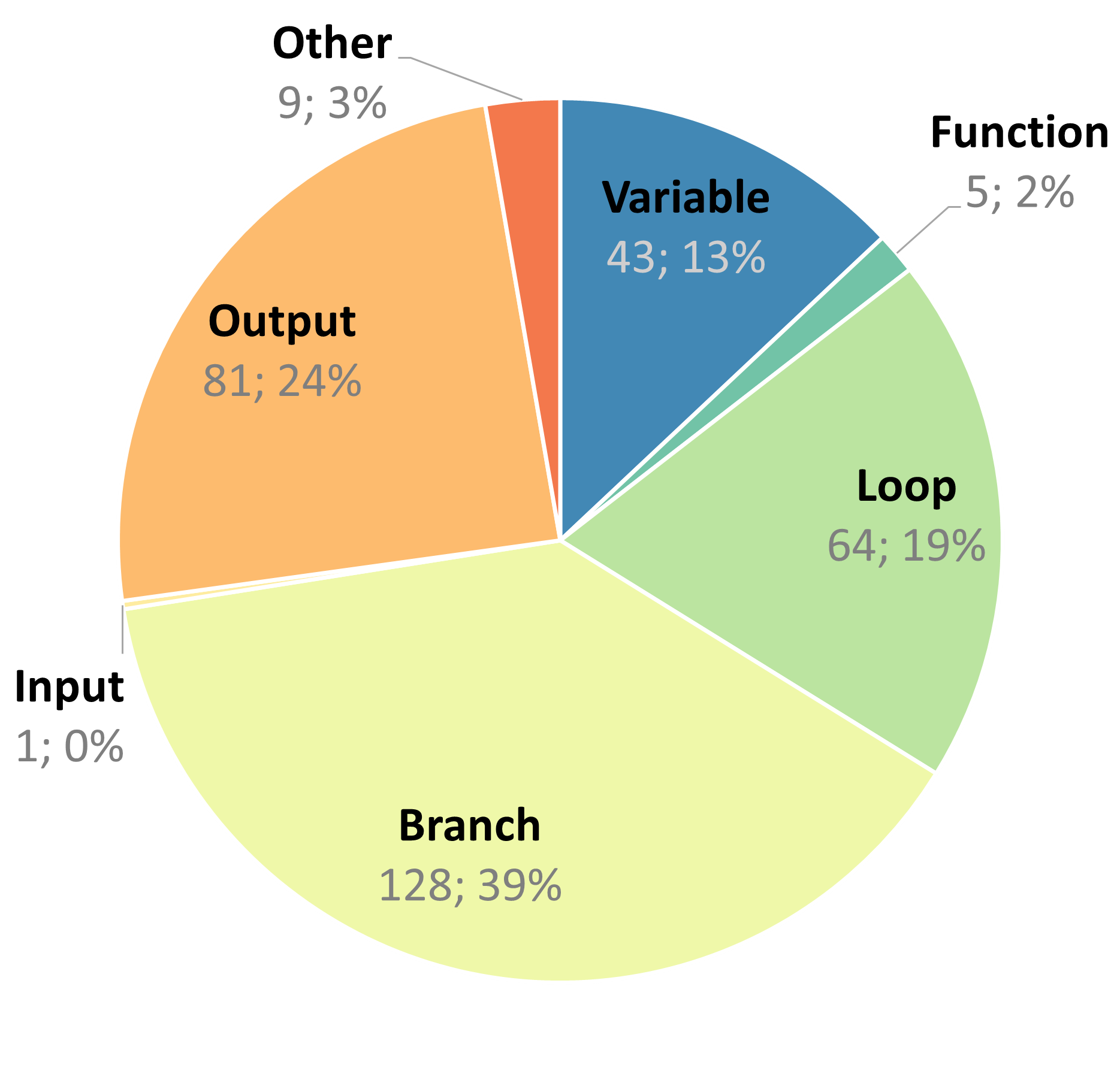}
         \includegraphics[width=0.45\textwidth]{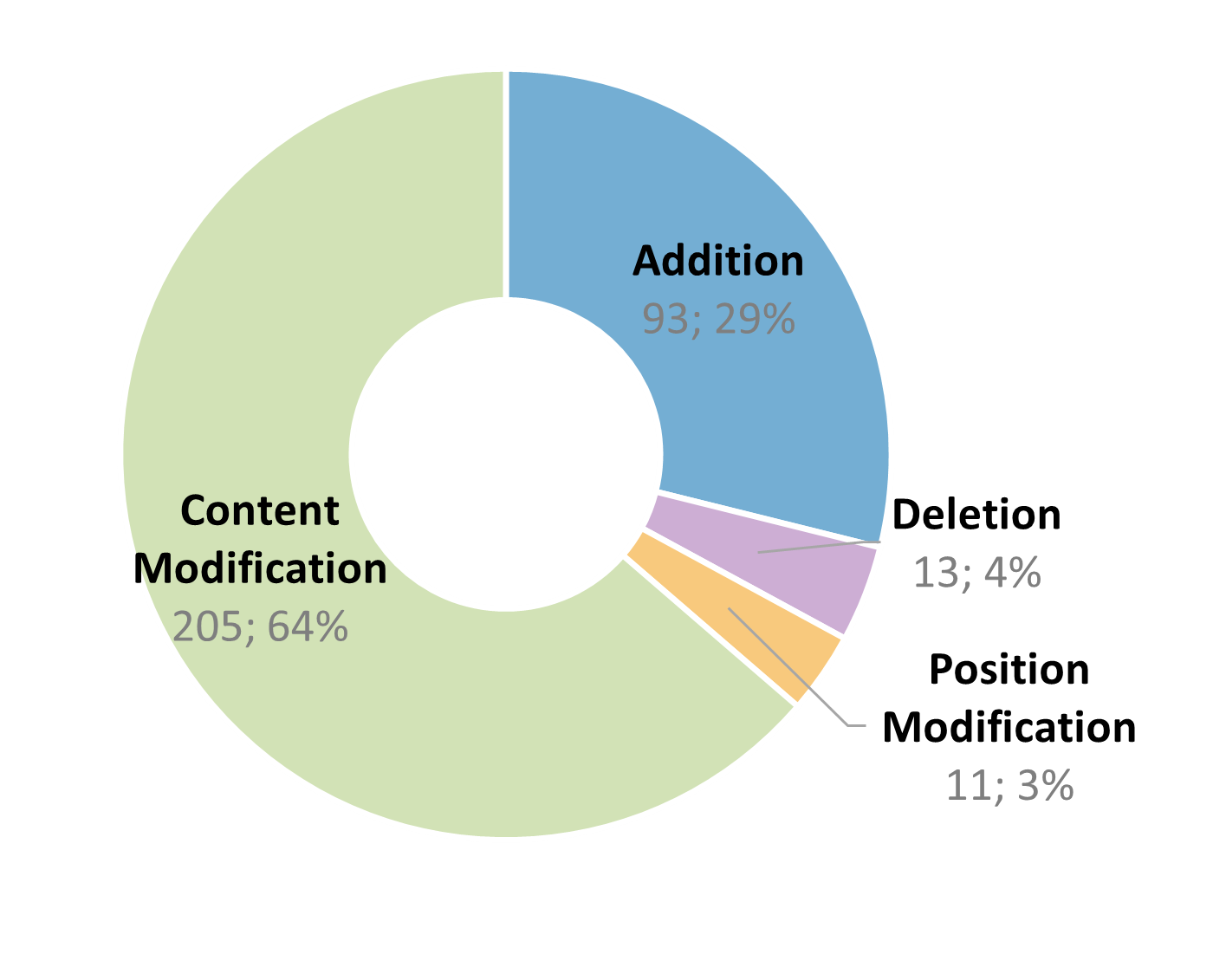}
         \caption{\textit{Prob.} 3}
         \label{fig:problem3}
     \end{subfigure}
     \begin{subfigure}[b]{0.45\textwidth}
        \setlength{\abovecaptionskip}{0cm}
         \centering
         \includegraphics[width=0.45\textwidth]{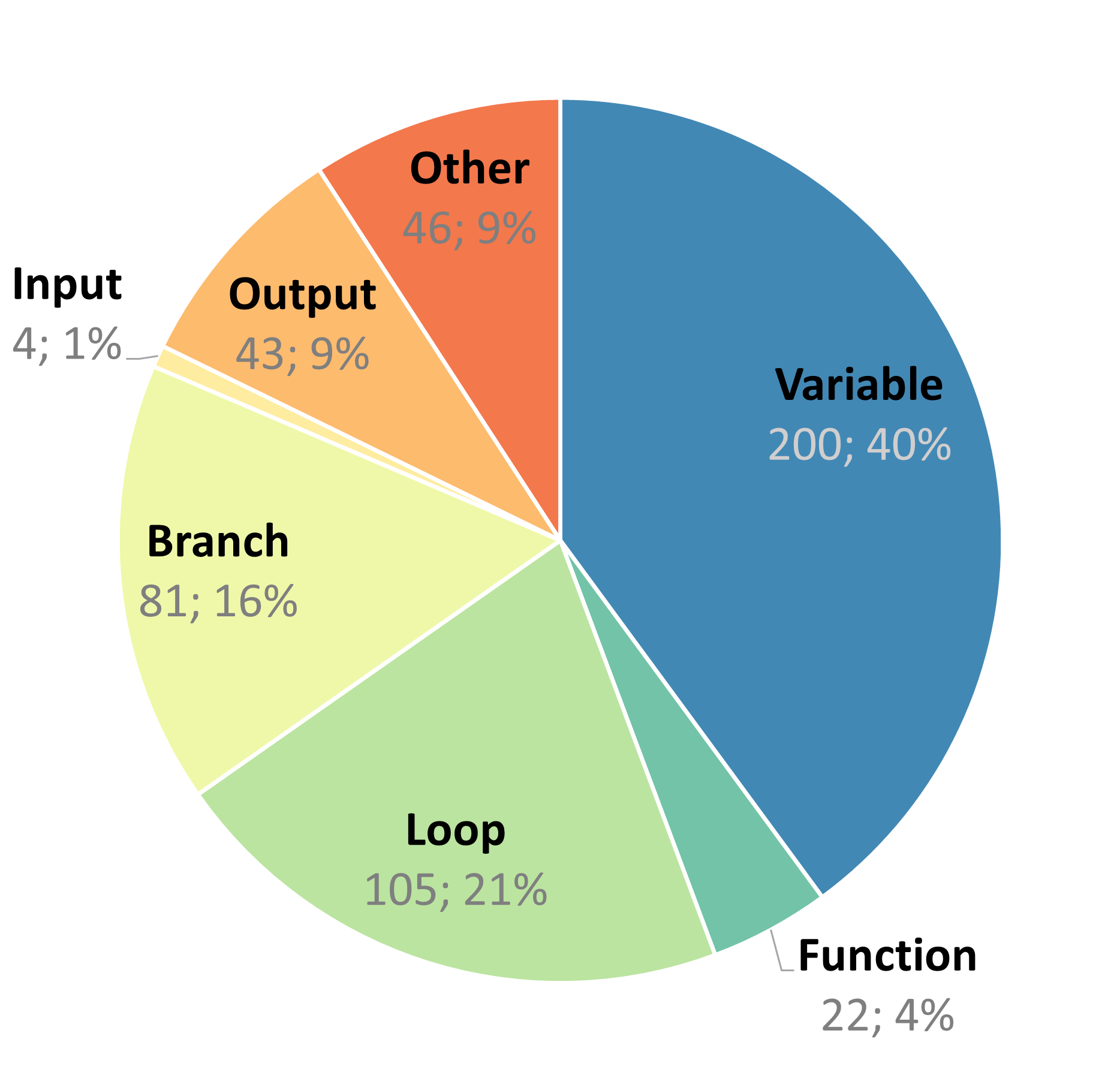}
         \includegraphics[width=0.45\textwidth]{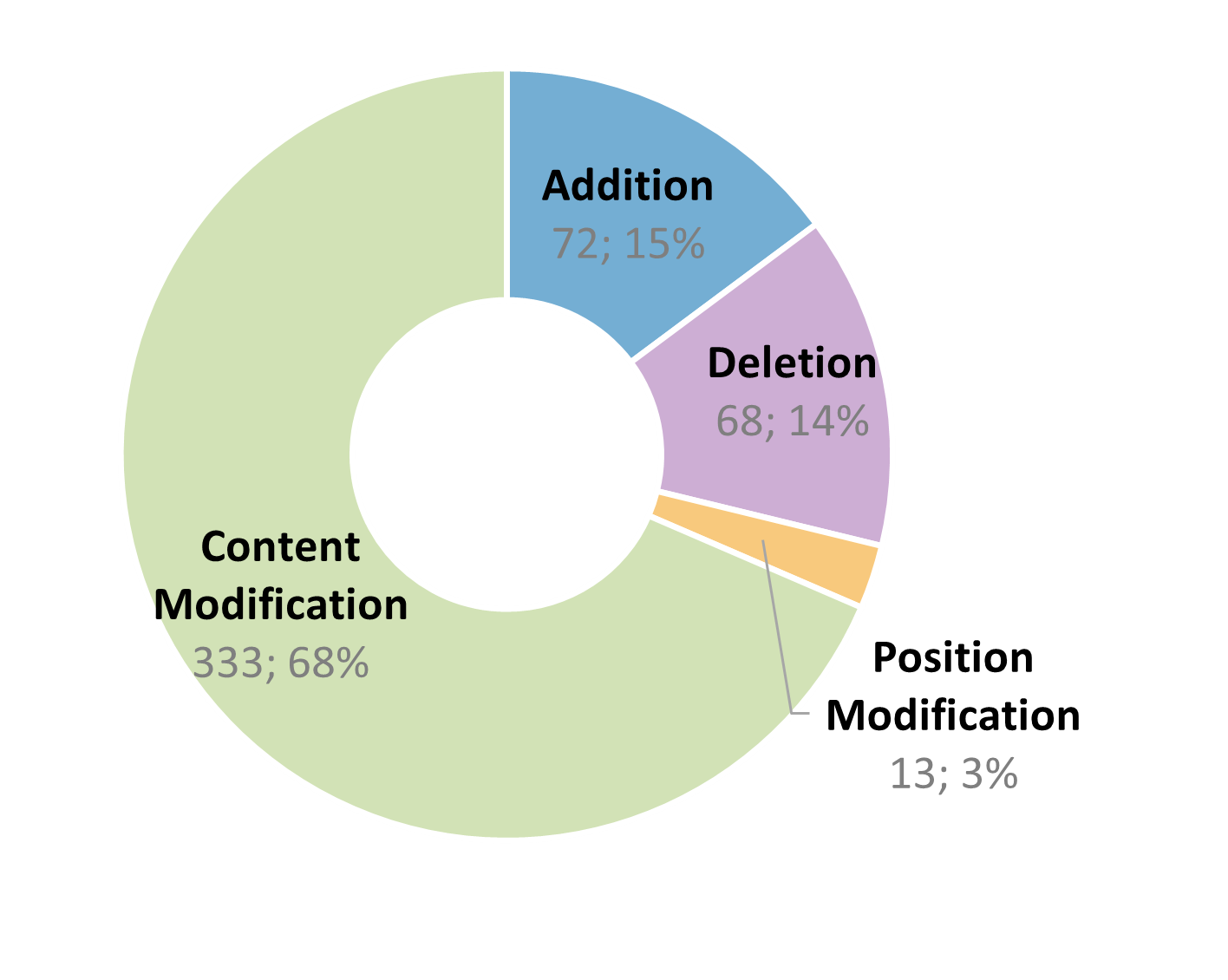}
         \caption{\textit{Prob.} 4}
         \label{fig:problem4}
     \end{subfigure}
        \caption{Distribution of bug types and repair types for \textit{Prob.}  1-4.}
        \label{fig:bug type distribution}
        \vspace{-0.3cm}
\end{figure}

\noindent \textbf{Distribution of the bug types and repair types.}
Since each problem is not characterized in the same way, the corresponding distribution of bug types and repair types varies. We make a further analysis and present the distribution in Figure \ref{fig:bug type distribution}.
\ding{182} \textit{Prob.} 1 requires generating the full permutation of the given number. 
109 (58\%) of the bugs are about outputs, with the majority (100 out of 109) being format errors. We can deduce that most students knew how to solve the problem but overlooked the specific output formatting requirements mentioned in the problem description.
\ding{183} \textit{Prob.} 2 requires calculating the result of an input expression. Bugs in this problem primarily involve variables and branches, with additional issues encountered during loop usage. The majority of branch-related bugs consist of if statement condition errors (31 out of 73) and missing branches (27 out of 73). And the variable-related bugs mostly involve incorrect assignments (65 out of 78). These patterns suggest that the main logic in most of the buggy code is incorrect. Significantly, this problem has the highest proportion of related bugs, indicating that extensive modifications are necessary to fix the code.
\ding{184} \textit{Prob.} 3 involves converting between decimal form and scientific notation. Multiple conversion and output logic need to be considered due to the range of possible scenarios. Similar to \textit{Prob.} 2, the majority of bugs are related to branches, with if statement condition errors and missing branches being the top issues. Bugs related to outputs and loops are also significant, but different from \textit{Prob.} 1, most output bugs involve incorrect content rather than formatting errors. In terms of bug repairs, the proportion of statement addition is the highest among all the problems. This suggests that there is significant missing logic in the buggy code, indicating that students may have overlooked specific situations.
\ding{185} In contrast to the previous 3 problems that mainly deal with one-dimensional data such as numbers or characters,
\textit{Prob.} 4 involves a plane coordinate system, which requires the calculation of the maximum number of line segments of a continuous line segment.
A majority of students used a \textit{struct} type to store the horizontal and vertical coordinates of the input line segments to preserve the correspondence.
The use of \textit{struct} involves multiple operations related to variables and requires attention to some rules and details, which may be easily ignored by students due to the lack of proficiency. Therefore, the proportion of bugs about variables is the highest in this problem.
Complex problem description and implementation logic may bring challenges to repair.

\begin{figure}[t]
    \setlength{\abovecaptionskip}{0.1cm}
     \centering
     \begin{subfigure}[b]{1.0\textwidth}
        \setlength{\abovecaptionskip}{0.1cm}
        \includegraphics[width=1.0\linewidth]{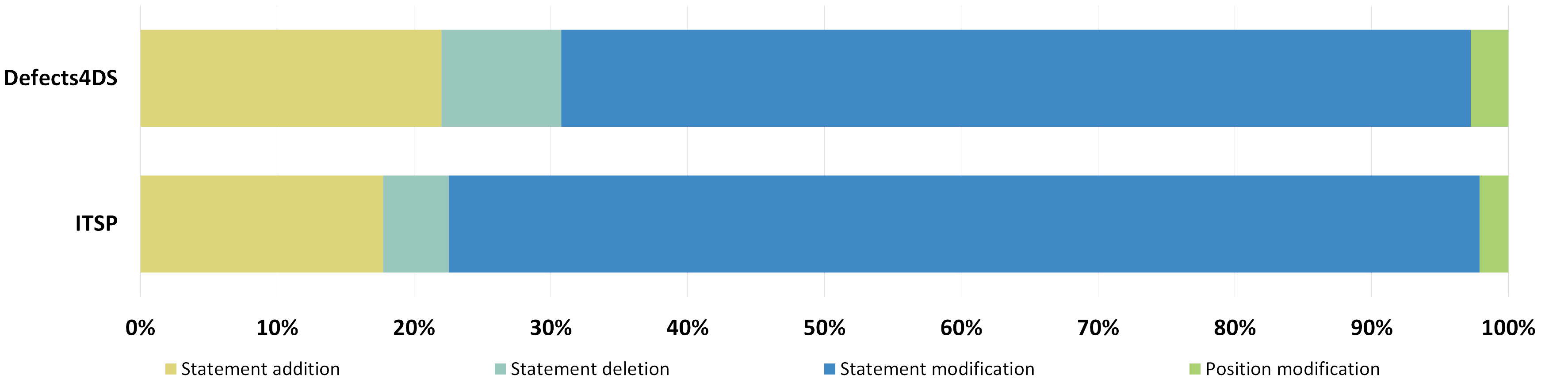}
        \caption{Comparison of the distribution of repair types.}
        \label{fig:itspVSds_repair}
     \end{subfigure}
     \begin{subfigure}[b]{1.0\textwidth}
        \setlength{\abovecaptionskip}{0.1cm}
        \includegraphics[width=1.0\linewidth]{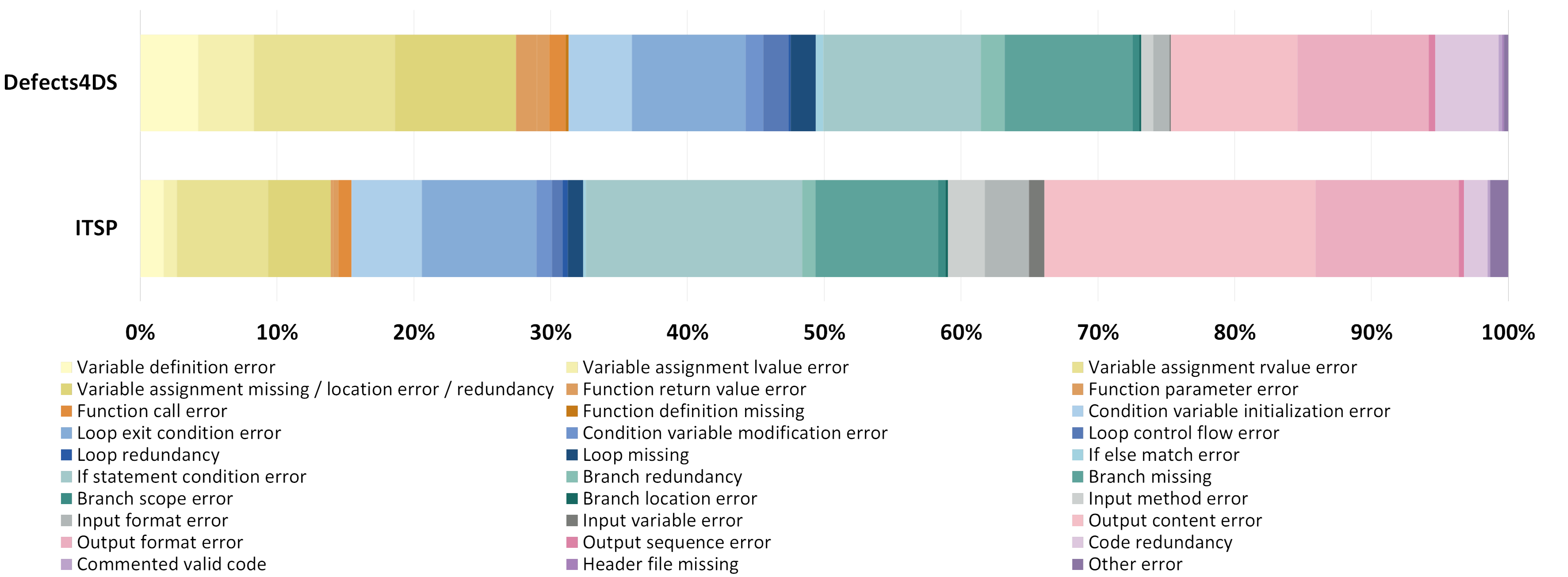}
        \caption{Comparison of the distribution of bug types.}
        \label{fig:itspVSds_bug}
     \end{subfigure}
     \caption{Comparison of the distribution of repair types and bug types in the annotated results between \ourdata{} and ITSP. To provide a more visually intuitive representation of the bug types, we utilize the same color scheme for different subcategories within the same main category.}
        \label{fig:itspVSds}
        \vspace{-0.4cm}
\end{figure}

\noindent \textbf{Comparison with ITSP.}
For the distribution of the number of bugs, from Table \ref{tab:number of bugs} we can observe that 65.4\% of the programs in ITSP contain only one bug, and the codes with more than three bugs constitute only 3.8\%. Regarding related bugs, the overall proportion in ITSP is 12.7\%, which is approximately one-fifth of the proportion in \ourdata{}.
Figure \ref{fig:itspVSds} showcases the comparison of repair types and bug types distribution between \ourdata{} and ITSP.
From Figure \ref{fig:itspVSds_repair}, it can be noted that the overall distribution of repair types in both datasets shows a similar pattern. However, \ourdata{} exhibits a higher proportion of statement additions and deletions, suggesting a greater extent of code changes required to repair the buggy code compared to ITSP.
Figure \ref{fig:itspVSds_bug} shows the distribution of bug types. It is obvious that the proportion of variable-related bugs in \ourdata{} far exceeds that in ITSP, almost doubling the rate. Variables are one of the most fundamental elements in code, and the usage of them permeates throughout. A small alteration in one place may potentially lead to significant changes in the final outcome. Besides, the output-related bugs constitute the largest proportion in ITSP, indicating that the majority of the computational logic in the code preceding them is correct. Furthermore, we also notice that \ourdata{} includes several bug types rarely or never found in ITSP, such as function return value error, loop control flow error, loop missing, if else match error, \textit{etc}. 

Based on the analysis above, and coupled with the cases involving complex grammatical components shown in Table \ref{tab:programming problems}, we believe the proposal of \ourdata{} contributes to the diversity and balance of bug types and comprehensive evaluation of APR methods.

\noindent \textbf{Challenges.}
Based on the above analysis, we outline the challenges faced by repairing the advanced student assignments as follows:

\noindent \textit{(1) Hard to locate errors}: The assignments often have larger code sizes and more syntax grammatical components, making them difficult to comprehend. Additionally, the presence of one or more semantic errors further complicates bug location.

\noindent \textit{(2) Hard to repair}: To fix the bugs, it is necessary to add or modify statements based on syntactic and semantic contextual information. Additionally, when multiple bugs are present, it is important to consider the correlation and mutual influence between them during the repair.

\noindent \textit{(3) Multi-source information exists}: Multiple sources of information are available during the repair, such as assignment descriptions, input/output format, example IOs, buggy code, passing conditions, peer solutions, \textit{etc}. While having a wealth of information provides more opportunities for repair, it can also create confusion. Extracting the most effective information from these sources is still a challenge.

\vspace{1mm}
\begin{mdframed}[linecolor=gray,roundcorner=12pt,backgroundcolor=gray!15,linewidth=3pt,innerleftmargin=2pt, leftmargin=0cm,rightmargin=0cm,topline=false,bottomline=false,rightline = false]
  \textbf{Finding}: It is difficult to identify and repair bugs in advanced student assignments due to the complex programs and the multiple related semantic bugs. The process is further complicated by diverse information sources, necessitating efficient information extraction methods. Thus, the proposal of \ourdata{} contributes to the diversity and balance of bug types and comprehensive evaluation of APR methods.
\end{mdframed}
\vspace{1mm}

\section{Approach}

\subsection{Overview}

\begin{figure}[t]
    \centering
    \setlength{\abovecaptionskip}{0.1cm}
    \includegraphics[width=1.0\linewidth]{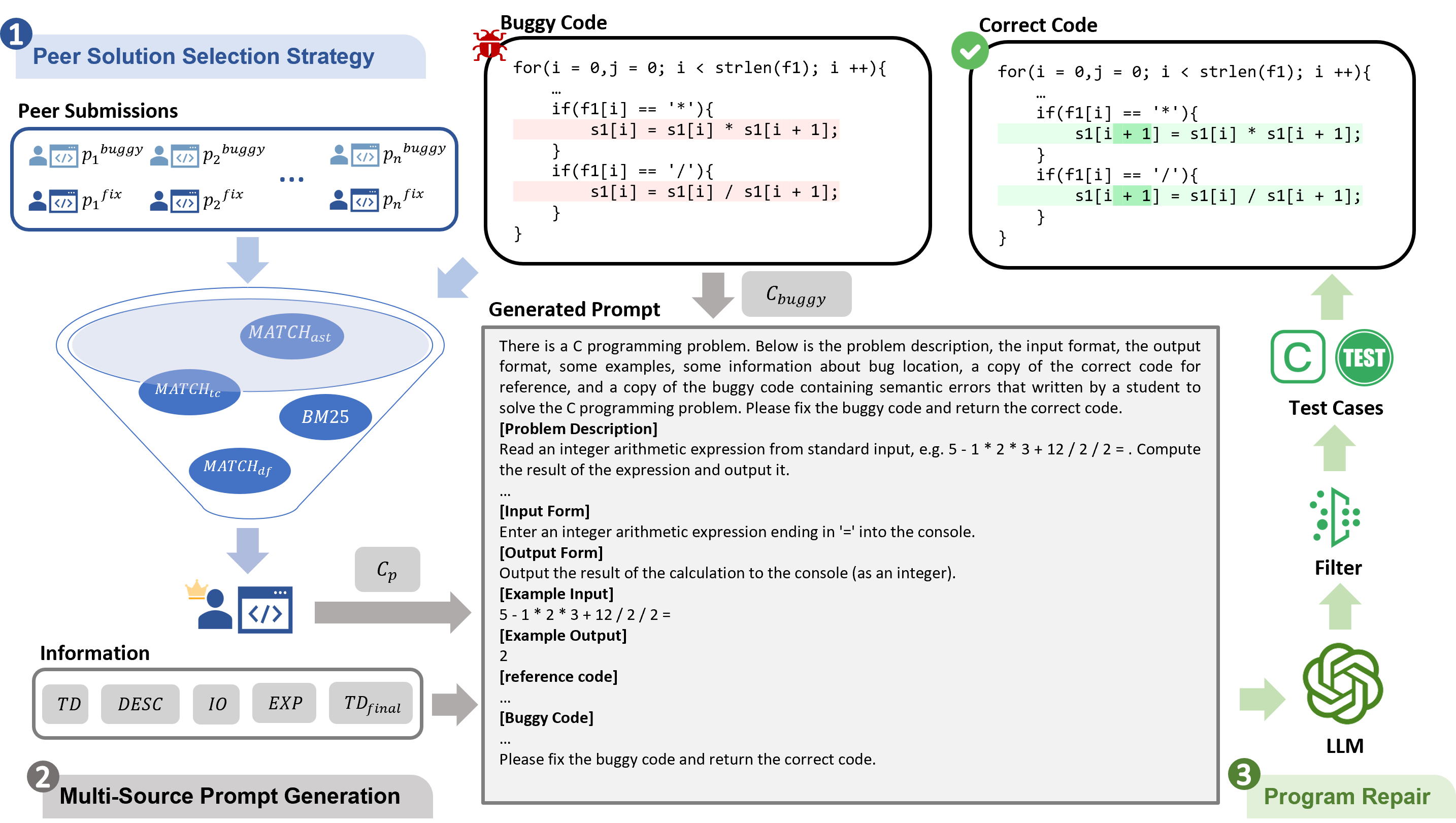}
    \caption{The architecture of \ourmodel{}.}
    \label{fig:architecture}
    \vspace{-0.3cm}
\end{figure}

To address the above challenges, we propose a novel repair framework \ourmodel{} for higher-level student assignment bug fixing. Figure \ref{fig:architecture} shows the architecture of our approach. To empower the LLM to fix the bugs in higher-level student programming assignments, we divide the procedure into three stages: Peer Solution Selection, Multi-Source Prompt Generation, and Program Repair.
During the peer solution selection stage, our model takes the buggy code $C_{buggy}$ as input and performs a comprehensive search to identify the most closely related solution $C_p$ from the peer submissions $P = \{(p_1^{buggy},p_1^{fix}), (p_2^{buggy},p_2^{fix}), ..., (p_n^{buggy},p_n^{fix})\}$, where $p_i^{fix}$ denotes the $i$-\textit{th} peer solution, and $p_i^{buggy}$ denotes its corresponding buggy code. 
Following that, in the prompt generation stage, our model combines multiple sources of information to craft a prompt $\mathcal{P}$ for the LLM, including the reference peer solution obtained in the previous step, program description, IO-related information, buggy code, \textit{etc.} 
Finally, in the program repair stage, the prompt generated by the previous stage is fed to the LLM, which then produces code $C_{fix}$ that is subsequently evaluated for correctness.

\subsection{Peer Solution Selection Strategy}\label{sec:Peer_select}
A major feature of the student programming assignments datasets that differs from other datasets is the availability of a substantial number of peer solutions that serve as valuable references. An effective reference code may improve the probability of correctly repairing code, while a poor reference code may impede the model's comprehension and generation, resulting in counterproductive outcomes. We argue that a valid reference code should closely resemble the buggy code in terms of the implementation details (lexical and syntactic). Besides, its corresponding buggy code has a similar error pattern (semantic) to the buggy code to be fixed. Therefore, we evaluate the similarity between a peer solution to the buggy code based on four aspects, which are as follows:

\noindent \textbf{$\textsc{match}_{tc}$: Test cases pass match score} (buggy-buggy). If two codes pass/fail the test cases in the same way, they probably have similar errors thus their repair methods may also be similar. This match score also reflects their semantic similarity from the perspective of execution status. This score is calculated by dividing twice the number of test cases that both codes can pass by the sum of the number of test cases that can be passed respectively.

\noindent \textbf{$\textsc{match}_{df}$: Data-flow match score} (buggy-correct). This score is further used to measure the semantic match of the buggy code and the correct peer code. The data flow score is calculated following \citet{ren2020codebleu}, which measures the proportion of the matched data flows.

\noindent \textbf{$\textsc{match}_{ast}$: AST match score} (buggy-correct). The abstract syntax tree (AST) is widely used to represent the syntactic structure of code. 
We obtain this match score by comparing the sub-trees of both codes to measure their similarity in terms of the syntactic structure.

\noindent \textbf{$\textsc{BM25}_{anon}$: Anonymous BM25 score} (buggy-correct). BM25 \cite{trotman2014bm25} is a metric that assesses the lexical relevance of documents.
We first calculate the BM25 score after variable anonymization by replacing variable names with generic names (v1, v2), alleviating the impact of different naming conventions, and then normalize the scores of each $C_{buggy}$ to be between 0 and 1.

Drawing upon the aforementioned aspects, we introduce a novel metric named Peer Solution Match ($\mathbf{PSM}$) score. This score is determined through a weighted sum of the above individual scores, enabling us to identify the most related peer solution. Notably, higher scores exemplify a stronger level of relevance.

\begin{equation}
\begin{split}
    \mathbf{PSM} &= \alpha \cdot \textsc{match}_{tc} + \beta \cdot \textsc{match}_{df} + \gamma \cdot \textsc{match}_{ast}  + \delta \cdot \textsc{BM25}_{anon} \\
    C_p &= \mathrm{argmax}_{p_i^{fix}}(PSM(C_{buggy}, (p_i^{bug}, p_i^{fix})), i=1,2,...n
\label{con:score}
\end{split}
\end{equation}

\noindent where $\alpha, \beta, \gamma, \delta$ are the weights for each score term, $n$ is the number of the peers, and $C_p$ is the final selected peer solution, which has the highest $\mathbf{PSM}$ score with the buggy code $C_{buggy}$.

\subsection{Multi-Source Prompt Generation} \label{sec:prompt generation}

As shown in Figure \ref{fig:architecture}, the prompt consists of three parts.
First is the general introduction, which describes the task and the detailed information introduced in the following parts. Then, more detailed content is provided, including the buggy code to be fixed and supplementary bug details to aid in the repair process. Finally, we repeat the task description concisely to prevent the model from forgetting after reading a long paragraph of text.\footnote{The impact of different prompt designing strategies is discussed in Section \ref{prompt_design}} Specifically, the following information is included:

\noindent \textbf{Task Description $(\mathit{TD})$}: This is the natural language used in the introduction that describes the task and the detailed information introduced in the following parts.

\noindent \textbf{Problem Description $(\mathit{DESC})$}: This is a detailed explanation, describing the task and the expected results.

\noindent \textbf{Input/Output Format $(\mathit{IO})$}: This provides a standardized description of the program input and output, aiming to prevent students from making errors due to format, such as spaces or newlines. 

\noindent \textbf{Example Input/Output $(\mathit{EXP})$}: This is to help students understand the meaning of the problem through one or more specific examples, sometimes accompanied by natural language explanations, which will not appear in the test cases.

\noindent \textbf{Reference Code $(\mathit{C}_p)$}: This comes from the most closely related peer solution retrieved using our peer solution selection strategy. It is added here as a reference given that students may ask other classmates for help when encountering difficulties.

\noindent \textbf{Buggy Code $(\mathit{C_{buggy}})$}: This is the buggy code to be fixed.

\noindent \textbf{Final Task Description $(\mathit{TD_{final}})$}: This is the repeated task description used at the end of the prompt.

\noindent We represent the prompt used in \ourmodel{} as $\mathcal{P}_{\ourmodel{}} = \{\mathit{TD}, \mathit{DESC}, \mathit{IO}, \mathit{EXP}, \mathit{C}_p, \mathit{C}_{buggy}, \mathit{TD}_{final}\}$.

\subsection{Program Repair}\label{sec:repair}
In this stage, we use the generated prompt $\mathcal{P}_{\ourmodel{}}$ as input of LLM.
Our framework is compatible with LLMs, allowing for seamless integration with a wide range of LLM models. 
To assess the generalizability of our framework, we applied PaR to ChatGPT \cite{chatgpt}, StarCoder (15.5B) \cite{li2023starcoder}, and Code Llama (34B) \cite{roziere2023codellama}. 
Since the output of LLM $O_{LLM}$ is sometimes a mixture of natural language and code, we use a filter $\mathcal{F}$ to retain code $C_{fix}$ only, and then send the code to an evaluation system $\mathcal{E}$ to verify whether it can pass all the test cases. 
The process can be summarized as follows:
\begin{equation}
    \begin{split}
        C_{fix} &= \mathcal{F}(O_{LLM}(\mathcal{P}_{\ourmodel{}})) \\
        \mathcal{E}(C_{fix}) &= \begin{cases} 
      \text{True}, & \text{if }C_{fix}\text{ passes all five test cases} \\
      \text{False}, & \text{otherwise}
   \end{cases}
    \end{split}
\end{equation}

\section{Evaluation}

We address the following research questions in our evaluation:

\vspace{-0.15cm}
\begin{center}\small
\begin{tcolorbox}[colback=gray!10,
                  colframe=black,
                  arc=1mm, auto outer arc,
                  boxrule=0.5pt,
                 ]
\textit{\textbf{RQ1: Performance Comparison.} How does the performance of \ourmodel{} in repairing programming assignments compared to state-of-the-art techniques? }

\textit{\textbf{RQ2: Impact of Prompt Design Strategy.} What are the contributions of various prompt components?}

\textit{\textbf{RQ3: Performance in Repairing Various Types of Bugs.} How does the effectiveness of resolving different types of bugs under different prompt settings and backbone models?}

\end{tcolorbox}
\end{center}
\vspace{-0.15cm}

We first demonstrate the performance of our model by comparing it against the state-of-the-art APR approaches on both \ourdata{} and an introductory programming assignment dataset (ITSP). Subsequently, we investigate the influence of various prompt components. Finally, we conduct a detailed analysis of the performance of both our model and baseline models across different buggy scenarios, exploring possible reasons for the observed results.

\subsection{Compared Techniques}

Our baselines can be mainly categorized into two types: LLM-based approaches and symbolic approaches. 
According to prior work \cite{zhang2022repairing, fan2023automated}, it is common to use the prompt which contains the problem description, the buggy code and the example IO (if available) as the input of LLM to repair the bugs, and their evaluation results have shown great performance on the introductory programming assignment repair datasets. Therefore, we also employ such an approach to evaluate the effectiveness of advanced LLMs on our dataset. 
Specifically, we aim to select LLMs recognized for their capability in program repair and applicability to Defects4DS, considering their maximum input length limits and model capacity. Following small-scale preliminary testing, we finally chose InCoder \cite{fried2022incoder}, CodeGen \cite{nijkamp2022codegen}, StarCoder \cite{li2023starcoder}, Codex \cite{chen2021evaluatingCodex}, ChatGPT \cite{chatgpt} and Code Llama \cite{roziere2023codellama} as our baselines.
InCoder \cite{fried2022incoder} is a large generative code model that can infill arbitrary regions of code. CodeGen \cite{nijkamp2022codegen} is a decoder-only model for multi-turn program synthesis. StarCoder \cite{li2023starcoder} is trained on English corpus and more than 80 programming languages using multi-query attention and offers two usage methods: generation (StarCoderPlus$_{Generate}$) and fill-in-the-middle (StarCoderPlus$_{Infill}$). 
Codex \cite{chen2021evaluatingCodex} is designed to understand and generate code built upon the GPT-3 architecture, which has been fine-tuned specifically for programming tasks. 
ChatGPT \cite{chatgpt} is a widely used language model developed by OpenAI, capable of understanding and generation tasks for conversational interactions.
Furthermore, we also include the recently released Code Llama model \cite{roziere2023codellama}. Code Llama, proposed by Meta AI, is a family of large language models for code based on Llama 2 \cite{touvron2023llama2}. It exhibits zero-shot instruction following capability for various programming tasks, which excels in performance compared to other open-source models.
We compare against these LLM-based baseline results on \ourdata{} dataset.

For symbolic approaches, we compare against Verifix \cite{Ahmed2021VerifixVR}, which is for providing verified repair to students undertaking introductory programming assignments. Despite demonstrating impressive repair capabilities on the widely utilized introductory assignment dataset (ITSP \cite{yi2017itsp}), Verifix falls short when it comes to our \ourdata{} dataset, where the pass rate is zero.
To identify the cause of the failure, we conduct a thorough analysis and present the reasons as follows. Verifix uses an off-the-shelf SMT solver, specifically the Z3 solver \cite{de2008z3}, to determine the minimal repair. However, the theory solvers in Z3 are limited to linear arithmetic, fixed-sized bit-vectors, arrays, and tuples. Consequently, they are unable to handle more complex grammatical components such as struct, pointer, and multidimensional arrays. As discussed in Section \ref{sec:data_statistics}, these components are present in 263 out of 682 (38.6\%) of \ourdata{} programs, while they are absent in the ITSP used by Verifix. Furthermore, there is a higher occurrence of custom functions in \ourdata{} compared to ITSP. This discrepancy suggests a significant gap in the difficulty of repairs.
To ensure a valid comparison, we compare the performance of \ourmodel{} and Verifix on the ITSP dataset.

\subsection{Implementation} \label{sec:implementation}
Regarding the specific configuration and implementation of our \ourmodel{} framework, we set all the coefficients in the formula (\ref{con:score}) to 1/4 ($ \alpha, \beta, \gamma, \delta = 1/4$). $\textsc{match}_{tc}$ are calculated based on the execution results of our data, $\textsc{match}_{df}$, $\textsc{match}_{ast}$ and $\textsc{BM25}_{anon}$ are calculated following the work of \cite{ren2020codebleu, zhu2022xlcost-codebleu, trotman2014bm25}. 
We use prompt $\mathcal{P}_{\ourmodel{}}$ mentioned in Section \ref{sec:prompt generation} and employ Python to execute the repair functions of \ourmodel{}.
As mentioned in Section \ref{sec:repair}, we choose three LLMs as the backbone model for \ourmodel{}, i.e., ChatGPT \cite{chatgpt}, StarCoder \cite{li2023starcoder}, and Code Llama \cite{roziere2023codellama}. For ChatGPT, we utilize the ChatGPT API gpt-3.5-turbo-0613 model\footnote{https://platform.openai.com/docs/models/gpt-3-5} endpoint. As for StarCoder, we utilize the StarCoderPlus$_{infill}$ version, which has 15.5B parameters. Lastly, for Code Llama, we employ the instruct version with 34B parameters. 
The average length of $\mathcal{P}_{\ourmodel{}}$ is 1270 tokens and the max length is 3027 tokens, within the max tokens limit of all three models.
To obtain a varied collection of potential patches, we use a sampling temperature of 0.8. All the codes are compiled and run based on the C99 standard.

As for LLM baselines, following previous work \cite{fan2023automated,xia2023keep,joshi2023repair}, 
we generate a basic prompt $\mathcal{P}_{\text{basic}} = \{\mathit{TD}, \mathit{DESC}, \mathit{IO}, \mathit{EXP}, \mathit{C}_{buggy}, \mathit{TD}_{final}\}$, which includes the information provided in the programming assignments.
For Codex, Code Llama, and ChatGPT, which are good at few-shot learning or conversational interactions, we obtain the fixed code by directly feeding the prompt into the model, without providing the bug location.
However, for both the infilling and generative baseline models, the bug location is required to perform repairs. Specifically, for the infilling models (InCoder and StarCoderPlus$_{Infill}$), we replace the buggy line with a special token and let the model complete it. For the generative models (CodeGen and StarCoderPlus$_{Generate}$), we retain the code before the appearance of the first bug and the rest is to be generated by the model. 
The input has an average length of around 900 tokens and the max length is around 2400 tokens. Prompts exceeding the max length limit of baseline models will be truncated.

\subsection{Metrics}
We use the following two metrics for evaluation: 

\noindent \textbf{Number of Fixed Programs (Pass Rate)}: As the behavior of LLM is inherently stochastic, while dealing with each buggy program, we initiate the bug repair request five times, each time within a new conversation context. We then classify the result based on a majority vote following \citet{cao2023study}, whereby if at least three out of the five replies correctly repair the code, we label it as ``pass''. 
on and the overall pass rate based on this criterion.
We calculate the pass rate by dividing the number of passed submissions by the total number of submissions for each specific problem as well as the entire dataset.
    
\noindent \textbf{Number of Partial Repaired Programs (Partial Repair Rate)}: Following \cite{yi2017itsp}, a repair candidate $P'$ is considered as a partial repair if (1) all previously passing tests still pass with $P'$, and (2) at least one of previously failing tests passes with $P'$. If at least three out of the five replies meet the above requirements, we consider the buggy code to have been partial repaired. It is worth mentioning that the fixed programs are included in the partial repaired programs.

\section{Results}
\subsection{RQ1: Performance Comparison}
\subsubsection{Results on \ourdata{}}
We first compare the bug-fixing ability of \ourmodel{} and LLM baselines in our advanced student assignment dataset \ourdata{}. By default, \ourmodel{} utilize the prompt setting of $\mathcal{P}_{\ourmodel{}}$. The results can be found in the rows corresponding to \textbf{\ourmodel{}} in Table \ref{tab:performance of different LLMs}. Furthermore, considering that most of the baselines require and include the bug location information, we conducted further experiments of introducing the Bug Location (BL) information into our prompt. Specifically, we add the line number(s) where the bug occurs into the prompt, as presented in the \colorbox{lightgrey}{\textit{Bug Line}} part in the upper-left of Figure \ref{fig:bug_prompt_example}.
The corresponding results are presented in the rows belonging to \textbf{\ourmodel{} w/BL} in Table \ref{tab:performance of different LLMs}. 
As seen from the results, we observed that all the variations of \ourmodel{} and \ourmodel{} w/BL achieved excellent performance, surpassing other baseline models by a significant margin.
When utilizing our default prompt setting, \ourmodel{}-ChatGPT achieves the best performance with a pass rate of 37.97\%. \ourmodel{}-CodeLlama also demonstrates superior performance compared to other baselines. 
After introducing the Bug Location information, we observe different results. With the help of BL information, the performance of \ourmodel{}-CodeLlama was significantly improved, with the pass rate increasing from 32.99\% to 44.13\%, making it the highest among all the models. Conversely, the performance of \ourmodel{}-ChatGPT became worse, with the pass rate decreasing from 37.97\% to 35.92\%.
\ourmodel{}-StarCoderPlus also obtains considerable performance compared to the baseline results, particularly when compared to the powerful ChatGPT and its counterpart.
Moreover, it is worth mentioning that these variations excelled in different problem areas, with \ourmodel{}-CodeLlama performing best in \textit{Prob.} 2 and 3, while \ourmodel{}-ChatGPT excelled in \textit{Prob.} 1 and 4. 
The output format in Problem 1 and the variable usage in Problem 4 involve relatively detailed issues. Code Llama may perform less effectively in handling such details, but it demonstrates greater competitiveness in correcting the overall logic of the code. This will be further discussed in subsequent sections.

Among all the baselines, the best performance is exhibited by ChatGPT, achieving a pass rate of 24.19\%, with StarCoderPlus$_{infill}$ and Code Llama falling behind.
The remaining models all exhibit relatively poor performance, with a pass rate below 5\%.
When analyzing the results in terms of generate mode, it becomes apparent that the generative approaches exhibit the poorest performance. This could be attributed to their inability to obtain complete information from the original flawed code and make modifications before the first bug. Consequently, this mode is significantly limited in the information they can access, unlike the other two modes. 
Furthermore, the infill models are also subject to limitations in some scenarios. They can only generate code at marked positions, which prevents them from addressing issues that require changes in statement placement or adding code elsewhere.
On the contrary, models utilizing the prompt mode do not suffer from these limitations, as they can access and comprehend all the information from the buggy code and repair it by generating complete code snippets.
Regarding comparisons within each generate mode, models with a higher number of parameters consistently outperform others. This is closely tied to the extensive knowledge they acquired during the pre-training phase.

\begin{table}[t]
    \centering
    \setlength{\abovecaptionskip}{0.1cm}
    \caption{Comparison of different LLMs on \ourdata{}. The numbers inside the parentheses represent the pass rate.
    }
    \resizebox{1.0\linewidth}{!}{
    \begin{tabular}{cllllllll}
    \toprule
    & \multirow{2}{*}{\textbf{Generate Mode}} & \multirow{2}{*}{\textbf{Model}} & \multirow{2}{*}{\textbf{\# Params}} & \multicolumn{5}{c}{\textbf{\# Fixed Programs}} \\
    \cmidrule(lr){5-9}
     &  &  & & \textbf{Prob.1} & \textbf{Prob.2} & \textbf{Prob.3} & \textbf{Prob.4} & \textbf{Overall} \\
     
    \midrule
    \multirow{7}{*}{Baseline} & \multirow{2}{*}{Infill} & InCoder & 1B & 0 & 4 & 7 & 21 & 32 \percent{(4.69\%)}\\
    & & StarCoderPlus$_{Infill}$ & 15.5B & 0 & 17 & 25 & 64 & 106 \percent{(15.54\%)}\\
     \cmidrule(lr){2-9}
    & \multirow{2}{*}{Generative} & CodeGen-multi & 2B & 7 & 4 & 2 & 0 & 13 \percent{(1.91\%)}\\
     & & StarCoderPlus$_{Generate}$ &  15.5B & 0 & 6 & 3 & 20 & 29 \percent{(4.25\%)}\\
     \cmidrule(lr){2-9}
    & \multirow{3}{*}{Prompt} & Codex (code-davinci-edit-001) & 12B  & 0 & 1 & 3 & 24 & 28 \percent{(4.11\%)}\\
    &  & Code Llama & 34B & 19 & 8 & 6 & 30 & 63 \percent{(9.24\%)}\\
    &  & ChatGPT & unknown & 76 & 11 & 5 & 73 & 165  \percent{(24.19\%)} \\
    
    \midrule
    \multirow{2}{*}{\ourmodel{}} & \multirow{2}{*}{Prompt} & \ourmodel{}-CodeLlama & 34B & 74 & \textbf{33} & \textbf{53} & 65 & 225 \percent{(32.99\%)}\\
    & & \ourmodel{}-ChatGPT & unknown & \textbf{91} & 25 & 29 & \textbf{114} & \textbf{259} \percent{(37.97\%)}\\
    \midrule
    \multirow{3}{*}{\ourmodel{} w/BL} & Infill & \ourmodel{}-StarCoderPlus$_{Infill}$ & 15.5B & 68 & 16 & 27 & 70 & 181 \percent{(26.54\%)}\\
    \cmidrule(lr){2-9}
    & \multirow{2}{*}{Prompt}& \ourmodel{}-CodeLlama & 34B & \textbf{92} & \textbf{41} & \textbf{82} & 86 & \textbf{301} \percent{(44.13\%)}\\
    & & \ourmodel{}-ChatGPT & unknown & 90 & 17 & 36 & \textbf{102} & 245 \percent{(35.92\%)}\\
    \bottomrule
    \end{tabular}
     }
    \label{tab:performance of different LLMs}
    \vspace{-0.3cm}
\end{table}

\begin{table}[h]
    \centering\small
    \setlength{\abovecaptionskip}{0.1cm}
    \caption{Pass rate of Verifix and \ourmodel{} on ITSP dataset (lab 3-6) which contains 28 assignments and 341 incorrect programs. The results of Verifix are sourced from \citet{Ahmed2021VerifixVR}. The numbers inside the parentheses represent the number of fixed programs.}
    \begin{tabular}{llllll}
    \toprule
    \multirow{2}{*}{\textbf{Model}} & \multicolumn{5}{c}{\textbf{Pass Rate}} \\
    \cmidrule(lr){2-6}
     & \textbf{Lab-3} & \textbf{Lab-4} & \textbf{Lab-5} & \textbf{Lab-6} & \textbf{Overall}\\
    \midrule
     Verifix & \textbf{92.1\%} & \textbf{82.9\%} & 45.1\% & 21.5\% & 58.4\% \percent{(199)}\\
     \ourmodel{}-ChatGPT & 52.4\% \da{39.7} & 76.1\% \da{10.8} & \textbf{80.5\%} \ua{35.4} & \textbf{79.7\%} \ua{58.2} & \textbf{73.6\%} \percent{(251)}\ua{\textbf{15.2}}\\
     \ourmodel{}-CodeLlama & 54.0\% \da{38.1} & 59.0\%  \da{6.1} & 70.7\% \ua{25.6} & 57.0\% \ua{35.5} & 60.4\% \percent{(206)} \ua{2}   \\
    \bottomrule
    \end{tabular}
    \label{tab:verifix vs gpt}
    \vspace{-0.3cm}
\end{table}

\subsubsection{Results on ITSP}
Then we further compare \ourmodel{} with Verifix \cite{Ahmed2021VerifixVR} on ITSP \cite{yi2017itsp} dataset. The settings of \ourmodel{} are the same as in Section \ref{sec:implementation}, meaning that all the coefficients in Formula (\ref{con:score}) are set to 1/4 and prompt $\mathcal{P}_{\ourmodel{}}$ is used.
The results are shown in Table \ref{tab:verifix vs gpt}. Overall, \ourmodel{}-ChatGPT achieves a pass rate of 73.6\%, surpassing Verifix by 15.2\%. Although the performance of \ourmodel{}-CodeLlama is slightly inferior, it is still better than Verifix.
Verifix aligns the student assignment with a reference solution in terms of control flow, and automatically summarizes differences in data variables using predicates to establish relationships between variable names. Hence, as the lab number increases, the difficulty of the tasks gradually escalates, code length grows, control flow becomes more intricate, and the number of variables increases. Consequently, the pass rate of Verifix gradually decreases. However, the pass rate of \ourmodel{}-ChatGPT generally remains stable at 80\%, except for Lab-3, where it is lower at 52.4\%. This divergence can be attributed to the conciseness of the Lab-3 task description, as \ourmodel{} primarily relies on the information provided in the prompt for repair. If the information is too brief or lacks sufficient details, it is likely to hinder its repair ability.

\vspace{1mm}
\begin{mdframed}[linecolor=gray,roundcorner=12pt,backgroundcolor=gray!15,linewidth=3pt,innerleftmargin=2pt, leftmargin=0cm,rightmargin=0cm,topline=false,bottomline=false,rightline = false]
  \textbf{Answer to RQ1:} \ourmodel{} consistently outperforms both LLM and symbolic approaches on both \ourdata{} and ITSP datasets, demonstrating its ability to empower large language models to effectively solve both advanced and introductory programming assignments. Additionally, \ourmodel{} demonstrates strong generalization capabilities and performs well across different LLMs.
\end{mdframed}
\vspace{1mm}

\subsection{RQ2: Impact of Prompt Design Strategies}\label{prompt_design}

The design of prompts significantly influences the performance of the model, as a well-crafted prompt can unlock its full potential. Hence, in this section, our objective is to delve into the impact of various prompt components. Initially, we concentrate on assessing the effect of IO-related information. Additionally, considering that many of our compared baselines require knowledge of the bug-related information, like bug location, before performing a repair, it is important to include and discuss the impacts of this information.
Given that we have pinpointed the bug-related information in the programs of \ourdata{}, we explore whether incorporating explicit buggy information in the prompt aids the model in the repair process. This evaluation sheds light on the significance and necessity of identifying bug information ahead of time, providing valuable insights and guidance.

For the IO-related information, we mainly explore the impact of example input/output ($\mathit{EXP}$) and test cases. The results can be seen in the upper part of Table \ref{tab: type of information}. 
As seen from the results, since the formats of $\mathit{EXP}$ and test cases are similar, both comprising an input and its corresponding output, there is no obvious difference in the number of fixed programs. 
This suggests that the prompts have similar effects, and LLM can capture enough knowledge from $\mathit{EXP}$.
Providing prompts with test cases tends to have a slightly higher pass rate and partial repair rate, but it comes with some risks. Revealing test cases that assess program correctness to LLM might lead it to generate code specifically tailored to those cases.
Therefore, we choose to use $\mathit{EXP}$ to prompt the LLM in \ourmodel{}.

\begin{table}[t]
    \centering
    \setlength{\abovecaptionskip}{0.1cm}
    \caption{Results of using different prompt designing strategies of \ourmodel{}-ChatGPT and \ourmodel{}-CodeLLama.}
    \resizebox{1.0\linewidth}{!}{
    \begin{tabular}{llcccccc}
    \toprule
    \multirow{2}{*}{\textbf{Model}} & \multirow{2}{*}{\textbf{Prompts}} & \multicolumn{5}{c}{\textbf{\# Fixed Programs}} & \multirow{2}{*}{\textbf{\# Partial Repaired}}  \\
    \cmidrule(lr){3-7} 
    & & \textbf{Prob.1} & \textbf{Prob.2} & \textbf{Prob.3} & \textbf{Prob.4} & \textbf{Overall} & \textbf{Programs}\\
    \midrule
    
    \rowcolor{lightgrey}
    \multicolumn{8}{l}{\textit{Example IO and Test Cases (Based on: $\mathit{TD}$ + $\mathit{DESC}$ + $\mathit{IO}$ + $\mathit{C}_{buggy}$ + $\mathit{TD}_{final}$)}}\\
    \multirow{3}{*}{\ourmodel{}-ChatGPT} & $\mathit{EXP}$ & 76 & 11 & 5 & 73 & 165 \percent{(24.19\%)} & 225\\
    & Test Cases & 80 & 14 & 18 & 65 & 177 \percent{(25.95\%)} & 231 \\
    & $\mathit{EXP}$ + Test Cases & 78 & 15 & 17 & 62 & 172 \percent{(25.22\%)} & 241\\

    \cmidrule(lr){1-8}
    \multirow{3}{*}{\ourmodel{}-CodeLlama} & $\mathit{EXP}$ & 19 & 8 & 6 & 30 & 63 \percent{(9.24\%)} & 90\\
    & Test Cases & 27 & 6 & 8 & 29 & 70 \percent{(10.26\%)} & 99\\
    & $\mathit{EXP}$ + Test Cases & 30 & 4 & 8 & 30 & 72 \percent{(10.56\%)} & 102 \\
    
    \midrule
    \rowcolor{lightgrey}
    \multicolumn{8}{l}{\textit{Bug Information (Based on: $\mathcal{P}_{\text{basic}}$ = $\mathit{TD}$ + $\mathit{DESC}$ + $\mathit{IO}$ + $\mathit{EXP}$ + $\mathit{C}_{buggy}$ + $\mathit{TD}_{final}$)}}\\
    \multirow{4}{*}{\ourmodel{}-ChatGPT} & Bug Line & 84 & 12 & 17 & 75 & 188 \percent{(27.57\%)} & 243\\
    & (+) Bug Type & 82 & 17 & 17 & 73 & 189 \percent{(27.71\%)} & 248\\
    & (+) Repair Type & 80 & 14 & 16 & 75 & 185 \percent{(27.12\%)} & 256\\
    & (+) Bug Correlation & 82 & 14 & 19 & 73 & 188 \percent{(27.57\%)} & 257\\

    \cmidrule(lr){1-8}
    \multirow{4}{*}{\ourmodel{}-CodeLlama} & Bug Line & 25 & 6 & 7 & 43 & 81 \percent{(11.88\%)} & 106 \\
    & (+) Bug Type & 67 & 10 & 9 & 32 & 118 \percent{(17.30\%)} & 151\\
    & (+) Repair Type & 58 & 12 & 7 &37 & 114 \percent{(16.72\%)}& 145\\
    & (+) Bug Correlation & 68 & 10 & 6 & 33 & 117 \percent{(17.16\%)} & 148\\
    
    \bottomrule
    \end{tabular}
    }
    \label{tab: type of information}
    \vspace{-0.3cm}
\end{table}

\begin{figure}[h]
    \centering
    \setlength{\abovecaptionskip}{0.1cm}
   \includegraphics[width=1.0\linewidth]{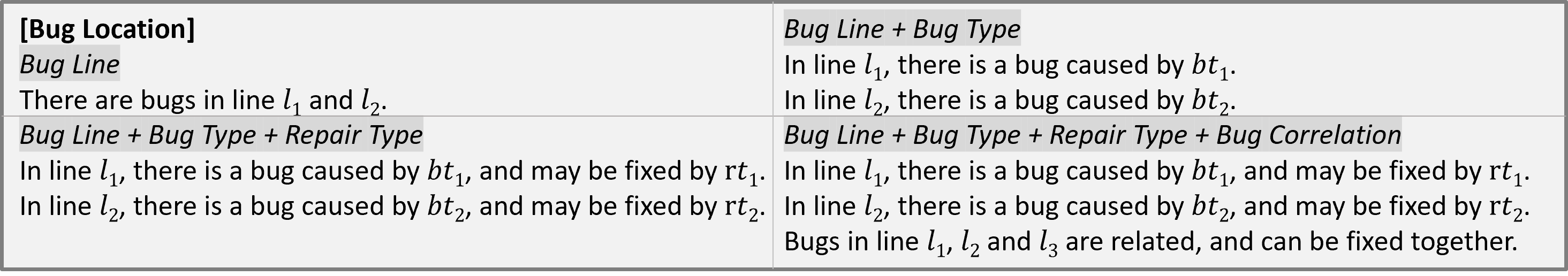}
    \caption{Example of the additional part of the prompt when considered the bug-related information. We use $\mathit{l}_{i}, \mathit{bt}_{i}$ and $\mathit{rt}_{i} \mathit{(i=1,2,3)}$ respectively to represent bug lines, bug types, and repair types. The texts in \colorbox{lightgrey}{\textit{Bug Line}} is utilized in \ourmodel{} w/BL in RQ1, while the remaining information is only used in RQ2.}
    \label{fig:bug_prompt_example}
    \vspace{-0.3cm}
\end{figure}

For the bug-related information, we mainly consider the following bug-related information as introduced in Section \ref{data_labeling}: Bug Line, Bug Type, Repair Type, and Bug Correlation. 
We add each of the above contents gradually into our basic prompt $\mathcal{P}_{\text{basic}}$, and the details can be seen in Figure \ref{fig:bug_prompt_example}.
The results are shown in the lower half of Table \ref{tab: type of information}. Given the conclusion we just reached, we add $\mathit{EXP}$ to the common portion of prompts. 
For \ourmodel{}-ChatGPT, we observe that the number of fixed programs is steady across these four prompts and slightly improved compared to those without bug information. The reason might be that ChatGPT inherently possesses strong comprehension and generation abilities. Thus, providing it with a bit of information related to bug localization, such as bug lines, allows it to uncover potential additional information on its own. 
While the trend presented by \ourmodel{}-CodeLlama is different. After adding the bug line information, the pass rate gains a slight improvement. Furthermore, there is a significant increase after adding bug type information. The pass rate and partial repair rate remain relatively stable with the addition of repair type and bug correlation information. Through analyzing the performance of each problem, We notice that the increase is mainly attributed to Problem 1. We deduce that Code Llama itself lacks certain bug location and judgment abilities. When provided with external hints and when the bug types are easy to understand and correct, it can achieve a certain level of improvement.
In comparison to Table \ref{tab:performance of different LLMs}, the absolute number of fixed programs and partial repaired programs is still quite low, which means that the increase brought by bug-related information is limited.
Considering the challenges involved in obtaining bug-related information in real-world scenarios and the marginal performance improvement observed, we are inclined to believe that the inclusion of bug-related information is not essential for \ourmodel{}. Consequently, while it is preferable to incorporate bug information, \ourmodel{} can still demonstrate effective performance even without it.

\vspace{1mm}
\begin{mdframed}[linecolor=gray,roundcorner=12pt,backgroundcolor=gray!15,linewidth=3pt,innerleftmargin=2pt, leftmargin=0cm,rightmargin=0cm,topline=false,bottomline=false,rightline = false]
  \textbf{Answer to RQ2:} 
  When designing the prompt, providing example input/outputs can sufficiently help LLMs understand the expected program behavior in terms of IO-related information. While bug-related information may offer only marginal benefits, \ourmodel{} can still demonstrate effective performance without it.
\end{mdframed}
\vspace{1mm}

\subsection{RQ3: Performance in Repairing Various Types of Bugs}
To comprehensively evaluate the performance across various buggy scenarios, we select four bug types that exhibit the highest occurrence frequencies in \ourdata{} based on our preliminary analysis (\textit{i.e.}, ``Variable'', ``Branch'', ``Loop'' and ``Output'') to investigate the effectiveness of repair under different prompt settings and backbone LLMs. 
We construct Venn diagrams to visually represent the results as depicted in Figure \ref{fig:venn_chatgpt}, \ref{fig:venn_codellama} and \ref{fig:venn gpt_cl}. The numbers in the diagrams represent the quantity of successfully repaired bugs, and the number inside the parentheses of the caption indicates the total number of bugs of this type.

\subsubsection{Different Prompt Settings}

\begin{figure}[t]
    \setlength{\abovecaptionskip}{0.1cm}
     \centering
     \begin{subfigure}[b]{0.20\textwidth}
         \centering
         \includegraphics[width=\textwidth]{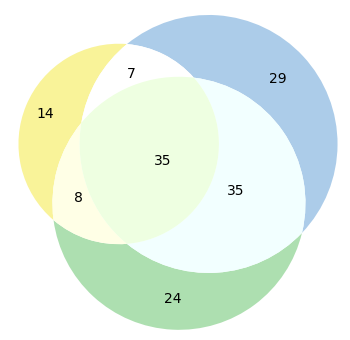}
         \caption{Variable (344)}
         \label{fig:variable_venn}
     \end{subfigure}
     \begin{subfigure}[b]{0.20\textwidth}
         \centering
         \includegraphics[width=\textwidth]{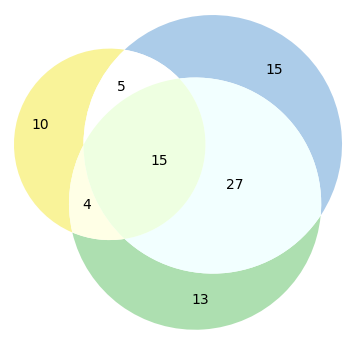}
         \caption{Branch (298)}
         \label{fig:branch_venn}
     \end{subfigure}
     \begin{subfigure}[b]{0.20\textwidth}
         \centering
         \includegraphics[width=\textwidth]{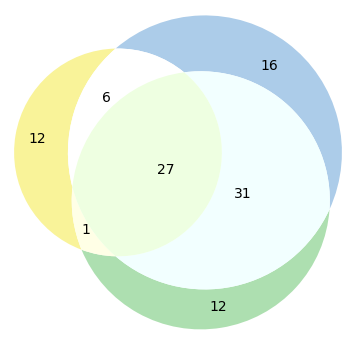}
         \caption{Loop (226)}
         \label{fig:loop_venn}
     \end{subfigure}
     \begin{subfigure}[b]{0.20\textwidth}
         \centering
         \includegraphics[width=\textwidth]{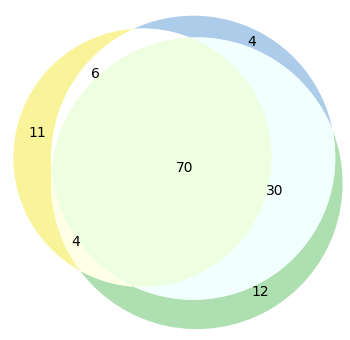}
         \caption{Output (242)}
         \label{fig:output_venn}
     \end{subfigure}
     \begin{subfigure}[b]{0.12\textwidth}
         \centering
         \includegraphics[width=\textwidth]{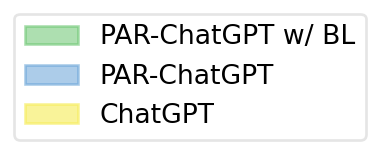}
     \end{subfigure}
    \caption{Results of employing ChatGPT as the backbone for repairing various types of bugs across various prompt settings.}
    \label{fig:venn_chatgpt}
    \vspace{-0.4cm}
\end{figure}

\begin{figure}[b]
    \setlength{\abovecaptionskip}{0.1cm}
     \centering
     \begin{subfigure}[b]{0.20\textwidth}
         \centering
         \includegraphics[width=\textwidth]{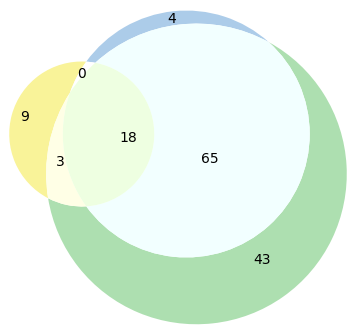}
         \caption{Variable (344)}
         \label{fig:variable_venn_cl}
     \end{subfigure}
     \begin{subfigure}[b]{0.20\textwidth}
         \centering
         \includegraphics[width=\textwidth]{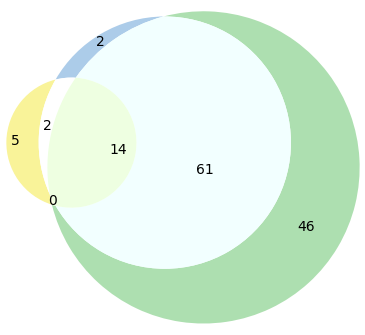}
         \caption{Branch (298)}
         \label{fig:branch_venn_cl}
     \end{subfigure}
     \begin{subfigure}[b]{0.20\textwidth}
         \centering
         \includegraphics[width=\textwidth]{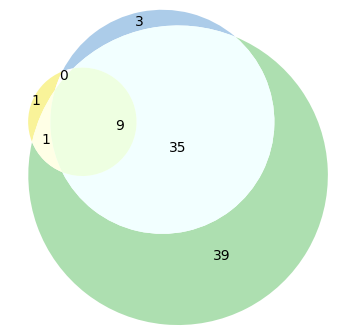}
         \caption{Loop (226)}
         \label{fig:loop_venn_cl}
     \end{subfigure}
     \begin{subfigure}[b]{0.20\textwidth}
         \centering
         \includegraphics[width=\textwidth]{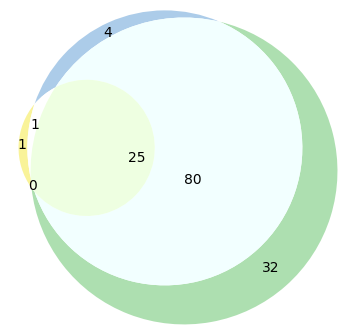}
         \caption{Output (242)}
         \label{fig:output_venn_cl}
     \end{subfigure}
     \begin{subfigure}[b]{0.12\textwidth}
         \centering
         \includegraphics[width=\textwidth]{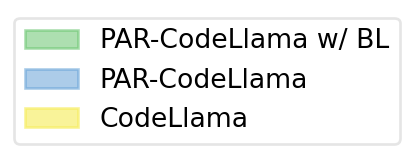}
     \end{subfigure}
    \caption{Results of employing Code Llama as the backbone for repairing various types of bugs across various prompt settings. }
    \label{fig:venn_codellama}
    \vspace{-0.4cm}
\end{figure}

For prompts, we mainly analyze the following settings: $\mathcal{P}_{\text{basic}}$, $\mathcal{P}_{\ourmodel{}}$, and $\mathcal{P}_{\ourmodel{}} \cup \mathit{BL}$. $\mathcal{P}_{\text{basic}}$ is used in the baseline as mentioned in Section \ref{sec:implementation}, which includes the information provided in the programming assignments. $\mathcal{P}_{\ourmodel{}}$ introduce the peer solution selected by our proposed strategy. $\mathcal{P}_{\ourmodel{}} \cup \mathit{BL}$ further incorporates the bug location information as in RQ1. The results obtained using ChatGPT and Code Llama as the backbone are illustrated in Figure \ref{fig:venn_chatgpt} and \ref{fig:venn_codellama}, respectively.

For ChatGPT-based models, as seen from Figure \ref{fig:output_venn}, the majority of the fixed \textit{output bugs} can be accomplished using any of the three prompt configurations. The rationale behind this is that \textit{output bugs} are relatively easier to repair among these four bug types because approximately half only involve formatting issues, such as missing spaces or incorrect line breaks, which do not involve deep-level code logic. 
However, the other three bug types, i.e., variable, branch, and loop bugs, involve more complex code logic and thus are more challenging to fix. Figures \ref{fig:variable_venn}, \ref{fig:branch_venn}, and \ref{fig:loop_venn} show that the bugs that can be repaired under all three prompts are in the minority, while \ourmodel{}-ChatGPT can repair more unique bugs under prompt $\mathcal{P}_{\ourmodel{}}$. Regarding the results of \ourmodel{}-ChatGPT w/BL, where the bug location information ($\mathit{BL}$) is introduced, \ourmodel{} can repair some new bugs but may fail to address bugs that have been previously fixed.
To sum up, the results suggest that for simple bugs, $\mathcal{P}_{\text{basic}}$ is also effective. $\mathcal{P}_{\ourmodel{}}$ is well-suited for resolving more complex issues, as it can fix a higher number of unique bugs compared to other prompts in this scenario. However, the introduction of $\mathit{BL}$ information does not always lead to beneficial outcomes.
For CodeLlama-based models, as seen from Figure \ref{fig:venn_codellama}, there is a notable disparity in the results compared to the results of ChatGPT. We can observe that only employing $\mathcal{P}_{\text{basic}}$, \textit{i.e.}, results of CodeLlama, performs poorly across all the bug types, demonstrating that CodeLlama is not good at locating and repairing the bug solely utilizing the information provided in the programming assignments. Utilizing $\mathcal{P}_{\ourmodel{}}$ as input leads to a significant performance improvement across all bug types. This indicates that $\mathcal{P}_{\ourmodel{}}$ provides valuable insights to guide Code Llama in effectively addressing the bugs. Moreover, it is important to highlight that the utilization of $\mathit{BL}$ information results in an additional performance improvement, contrasting with the behavior observed in ChatGPT. The results further confirm that CodeLlama continues to face challenges in locating bugs. However, when the bug location is explicitly provided, it can carry out bug fixes more effectively.

\subsubsection{Different Backbone Models} 
\begin{figure}[h]
    \setlength{\abovecaptionskip}{0.1cm}
     \centering
     \begin{subfigure}[b]{0.20\textwidth}
         \centering
         \includegraphics[width=\textwidth]{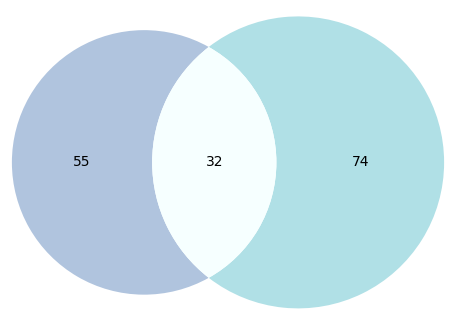}
         \caption{Variable (344)}
         \label{fig:baseline_variable_venn}
     \end{subfigure}
     \begin{subfigure}[b]{0.22\textwidth}
         \centering
         \includegraphics[width=\textwidth]{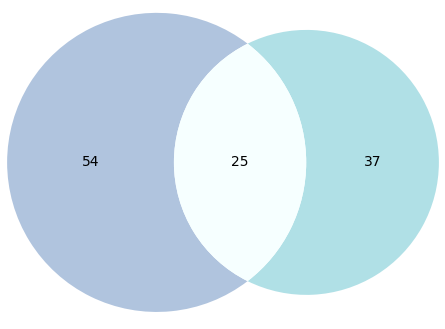}
         \caption{Branch (298)}
         \label{fig:baseline_branch_venn}
     \end{subfigure}
     \begin{subfigure}[b]{0.20\textwidth}
         \centering
         \includegraphics[width=\textwidth]{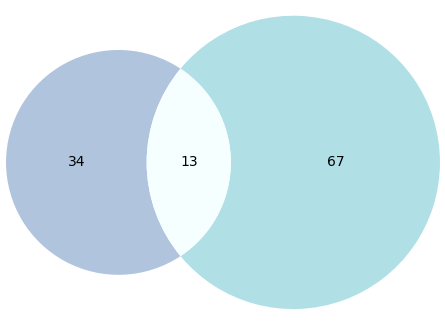}
         \caption{Loop (226)}
         \label{fig:baseline_loop_venn}
     \end{subfigure}
     \begin{subfigure}[b]{0.20\textwidth}
         \centering
         \includegraphics[width=\textwidth]{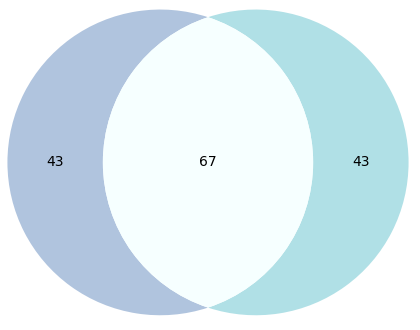}
         \caption{Output (242)}
         \label{fig:baseline_output_venn}
     \end{subfigure}
     \begin{subfigure}[b]{0.12\textwidth}
         \centering
         \includegraphics[width=\textwidth]{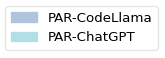}
     \end{subfigure}
    \caption{Bug fix Venn diagrams on \ourdata{} dataset under different models.}
    \label{fig:venn gpt_cl}
    \vspace{-0.3cm}
\end{figure}

Subsequently, we compare the performance of using two backbone models in repairing various types of bugs.
As depicted in Figure \ref{fig:venn gpt_cl}, the overlap between the \ourmodel{}-ChatGPT and \ourmodel{}-CodeLlama for resolving \textit{output bugs} is greater than for other types of bugs, demonstrating most of the \textit{output bugs} can be addressed by both of these two models. However, regarding other types of bugs, the overlap is much smaller, showcasing that there are distinct strengths and specializations in each model for handling different types of bugs. In particular, \ourmodel{}-ChatGPT excels in resolving \textit{Loop} and \textit{Variable} bugs, whereas \ourmodel{}-CodeLlama demonstrates superior performance in fixing \textit{Branch} bugs. These findings suggest that leveraging different backbone models for diverse bug types in practice may be beneficial, which will be explored in our future research.

\vspace{1mm}
\begin{mdframed}[linecolor=gray,roundcorner=12pt,backgroundcolor=gray!15,linewidth=3pt,innerleftmargin=2pt, leftmargin=0cm,rightmargin=0cm,topline=false,bottomline=false,rightline = false]
  \textbf{Answer to RQ3:} Compared to the basic prompt, $\ourmodel{}$ can repair more unique bugs and maintain its effectiveness across different types of bugs, especially when dealing with challenging bugs. Regarding different backbone models, each model has specific strengths and specializations for addressing various types of bugs. Additionally, providing bug location information is of greater importance for the Code Llama backbone. 
\end{mdframed}
\vspace{1mm}

\section{discussion}
\subsection{Impact of different peer solution selection strategies}

The previous results have shown that an effective reference code $\mathit{C}_p$ can provide valuable guidance, improving the repair performance substantially. Therefore, we evaluate the effectiveness of different selection strategies in this section, which can be divided into three types: lexical search, semantical search, and our proposed selection strategies with different parameter settings.
We use BM25 \cite{trotman2014bm25} algorithm for lexical search. We calculate the BM25 scores between $\mathit{C}_{buggy}$ and other buggy codes, and retrieve the code with the highest score as $\mathit{C}_p$.
For semantic search, we use two code semantic learning models. The first one is UniXcoder \cite{guo2022unixcoder}, which can retrieve codes with the same semantics from a collection of candidates in a zero-shot setting given a source code as the query.
NCC \cite{ben2018ncc} is an LLVM IR-based code representation model, which learns code semantic relying on an embedding space and graph representation of LLVM IR statements and their context.

For our selection strategy, except for the four aspects mentioned in Section \ref{sec:Peer_select}, we also explore the impact of considering bug types when searching for a similar peer solution, and the definition of this aspect is as follows: 

\noindent \textbf{$\textsc{match}_{bt}$: Bug type match score} (buggy-buggy). The match of the bug types of the two buggy codes (one representing the code to be fixed and the other representing the corresponding buggy code of the peer solution) provides insights into their semantic similarity. We calculate this score by dividing twice the number of overlapping bug types (both the buy type and the repair type are the same) by the sum of the number of bugs in both buggy codes.

\noindent After importing the $\textsc{match}_{bt}$, the Peer Solution Match ($\mathbf{PSM}$) score is computed as follows:
\begin{equation}
\begin{split}
    \mathbf{PSM} = \alpha \cdot \textsc{match}_{tc} + \beta \cdot \textsc{match}_{df} + \gamma \cdot \textsc{match}_{ast}  + \delta \cdot \textsc{BM25}_{anon} + \eta \cdot \textsc{match}_{bt} 
\label{con:score_new}
\end{split}
\end{equation}
where $\eta$ is the weight for the Bug-type match score.
Therefore, except for the default parameter setting ($ \alpha, \beta, \gamma, \delta = 1/4$), we experiment with a new parameter setting ($ \alpha, \beta, \gamma, \delta, \eta = 1/5$).
Besides, we also try another weight set, \textit{i.e.}, $\alpha=1/9$, $\beta=1/9$, $\gamma=1/9$, $\delta=1/3$, $\eta=1/3$, as we divide the five aspects into three categories and give each category equal weight: (1) lexical: $\textsc{BM25}_{anon}$; (2) syntactic: $\textsc{match}_{ast}$; (3) semantic: $\textsc{match}_{bt}$, $\textsc{match}_{tc}$, and $\textsc{match}_{df}$.

Results in Table \ref{tab:peer solution} demonstrate that our selection strategy surpasses all other search methods. Specifically, our default setting achieves the best result in terms of Partial Repair Rate. When introducing the bug types while selecting the peer solution, one more program is correctly fixed. Besides, when adjusting the weights for each aspect, the results also change slightly. 
This further suggests that our initial strategy embraces multiple evaluation metrics, enabling a comprehensive assessment of the code similarity from different angles. It is noteworthy that incorporating bug information does not explicitly enhance the performance, indicating that the bug information may not be crucial enough.
Among all the baseline strategies, UniXcoder outperforms the rest, demonstrating the effectiveness and significance of incorporating semantic information in the peer solution selection process.

\begin{table}[t]
    \centering
    \setlength{\abovecaptionskip}{0.1cm}
     \caption{Results of using different peer solution strategies when using ChatGPT as the backbone model. Other settings are the same with \ourmodel{}.}
    \label{tab:peer solution}
    \resizebox{1.0\linewidth}{!}{
    \begin{tabular}{lcccccc}
    \toprule
    \multirow{2}{*}{\textbf{Peer Solution Selection Strategy}} & \multicolumn{5}{c}{\textbf{\# Fixed Programs}} & \multirow{2}{*}{\textbf{\# Partial Repaired}}\\
    \cmidrule(lr){2-6} 
     & \textbf{Prob.1} & \textbf{Prob.2} & \textbf{Prob.3} & \textbf{Prob.4} & \textbf{Overall} & \textbf{Programs} \\
    \midrule
    \rowcolor{lightgrey}
    \multicolumn{7}{l}{\textit{Lexical Search}}\\
    $\textsc{BM25}$ & 87 & 22 & 30 & 85 & 224 \percent{(32.84\%)} & 300\\
    \midrule
    \rowcolor{lightgrey}
    \multicolumn{7}{l}{\textit{Semantic Search}}\\
    UniXcoder code-to-code search \cite{guo2022unixcoder} & 89 & 16 & 33 & 110 & 248 \percent{(36.36\%)} & 303\\
    NCC \cite{ben2018ncc} & 79 & 19 & 28 & 93 & 219 \percent{(32.11\%)} & 260\\
    \midrule
    \rowcolor{lightgrey}
    \multicolumn{7}{l}{\textit{Our Selection Strategy}}\\
    $\alpha=1/4$, $\beta=1/4$, $\gamma=1/4$, $\delta=1/4$ & 91  &  \textbf{25} & 29  & \textbf{114}  & 259  \percent{(37.97\%)} & \textbf{321} \\
    $\alpha=1/5$, $\beta=1/5$, $\gamma=1/5$, $\delta=1/5$, $\eta=1/5$ & 89 & 14 & \textbf{44} & 113 & \textbf{260 \percent{(38.12\%)}} & 307\\
    $\alpha=1/9$, $\beta=1/9$, $\gamma=1/9$, $\delta=1/3$, $\eta=1/3$ & \textbf{95} & 14 & 37 & 111 & 257 \percent{(37.68\%)} & 297\\
    \bottomrule
    \end{tabular}
    }
    \vspace{-0.3cm}
\end{table}

\subsection{Case Study/Qualitative Analysis}

In this section, we further assess how well the generated feedback supported students in understanding and resolving issues on their own, and whether the selected similar peer solutions can provide useful information for resolving similar issues. 

\subsubsection{Effectiveness in assisting students solving problems.}

\begin{figure}[t]
    \setlength{\abovecaptionskip}{0cm}
     \centering
     \begin{subfigure}[b]{0.43\textwidth}
         \centering
         \includegraphics[width=\textwidth]{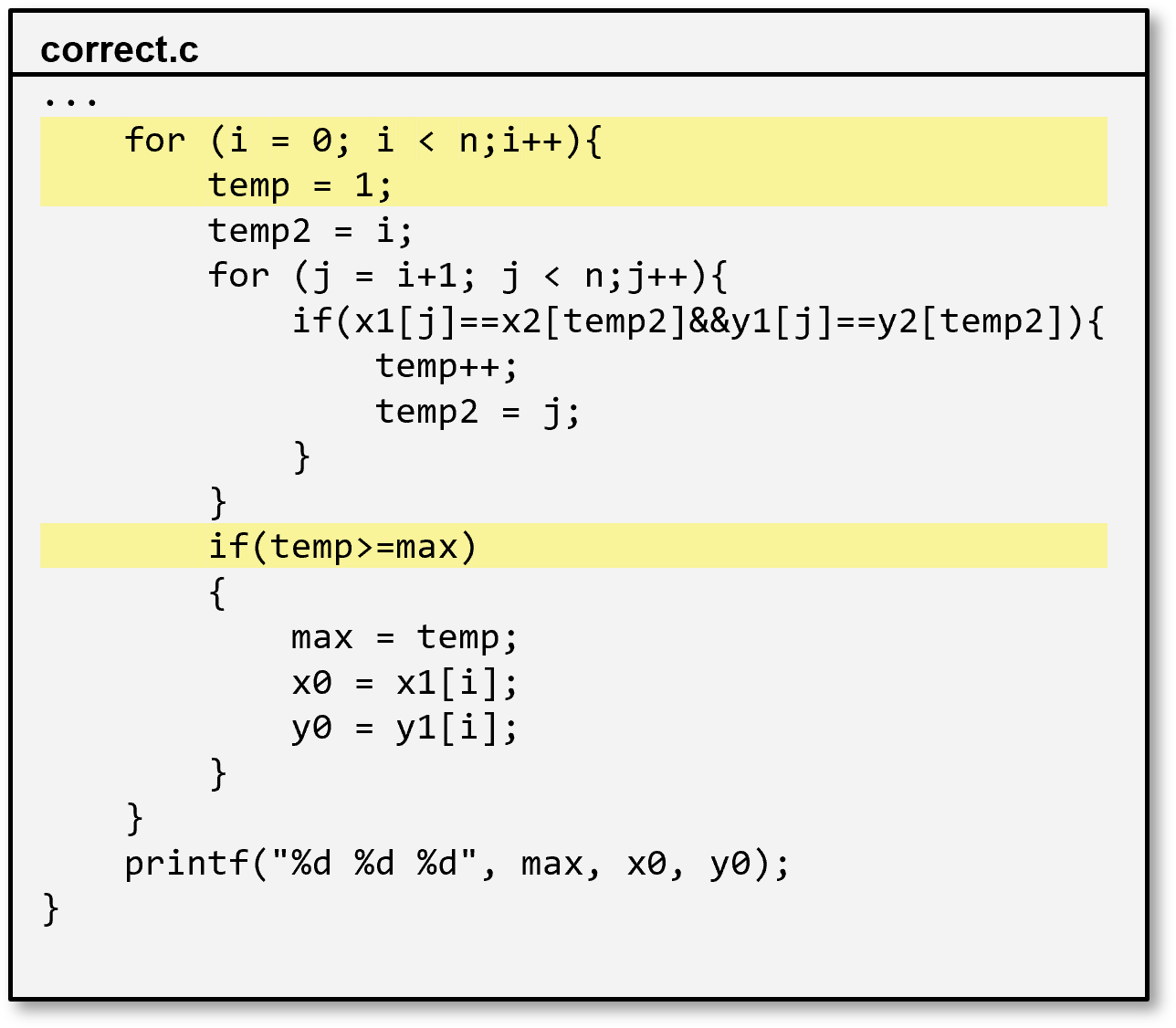}
     \end{subfigure}
     \begin{subfigure}[b]{0.43\textwidth}
         \centering
         \includegraphics[width=\textwidth]{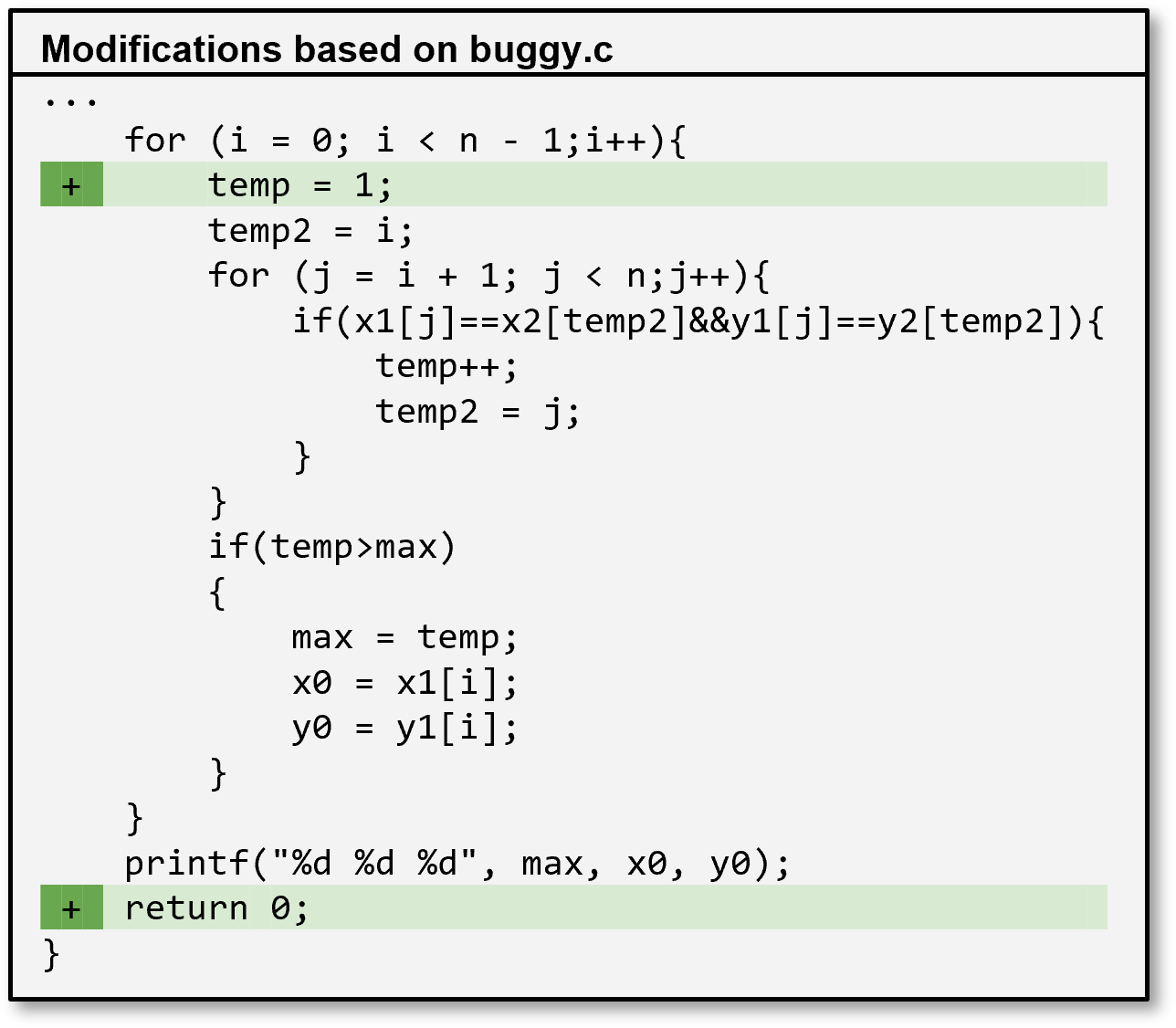}
     \end{subfigure}
    \caption{Bug fix example. The \textit{correct.c} code is the correct code submitted by the student, where the changed code is highlighted. The \textit{modifications based on buggy.c} is the fixed code generated by \ourmodel{}.}
    \label{fig: effectiveness of PAR}
    \vspace{-0.4cm}
\end{figure}

Figure \ref{fig: effectiveness of PAR} illustrates an example of a successful repair by \ourmodel{}. The critical error is the absence of reassigning the variable at the beginning of each loop iteration. The correct file submitted by the student made three changes to fix the bug, but two of them could not affect the output (different from the equivalent modifications mentioned in section \ref{data_labeling}). 
One is the change of loop exit condition. In this nested loop, the maximum value of variable $j$ is $n-1$, which implies that the maximum value of variable $i$ can only reach $n-2$. Therefore, there is no difference between $i<n$ and $i<n-1$ here.
The other is to replace $>=$ with $>$ in the if statement condition, which will not affect the result of the swapping operation at this point.
The modifications generated by \ourmodel{} show that it clearly discovered this fact and made relevant changes. This demonstrates that \ourmodel{} is able to provide effective assistance to students.

\begin{figure}[b]
    \setlength{\abovecaptionskip}{0cm}
     \centering
     \begin{subfigure}[b]{0.43\textwidth}
         \centering
         \includegraphics[width=\textwidth]{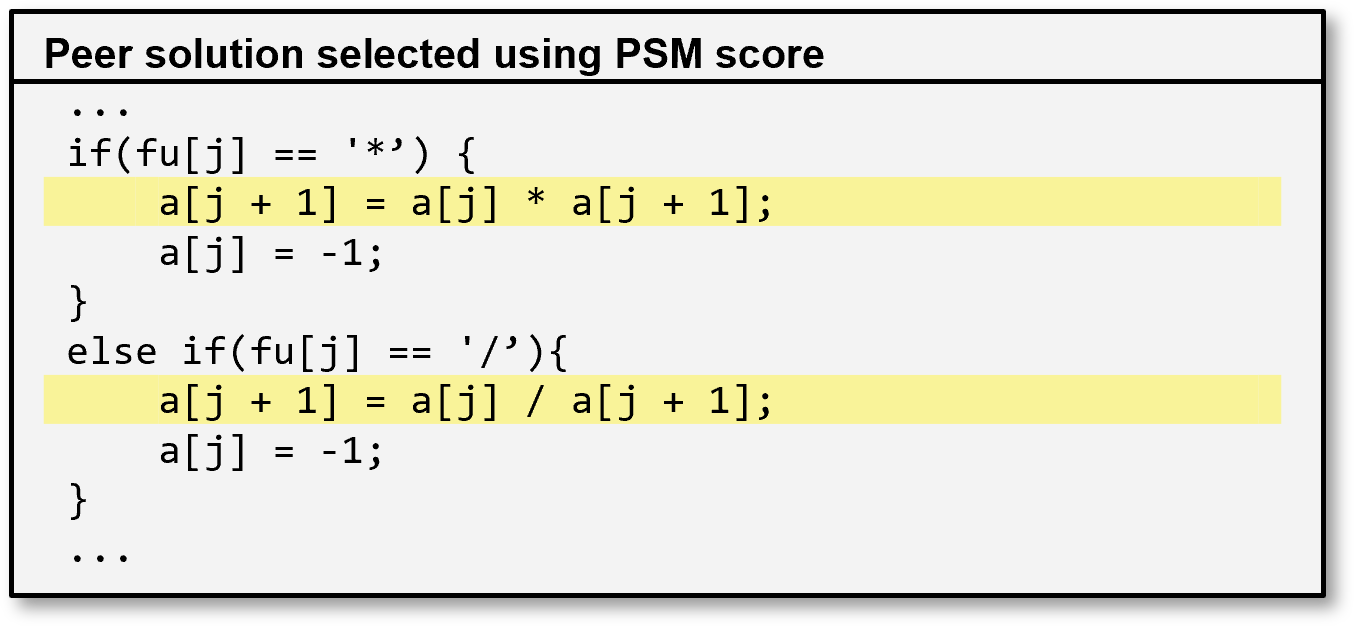}
     \end{subfigure}
     \begin{subfigure}[b]{0.43\textwidth}
         \centering
         \includegraphics[width=\textwidth]{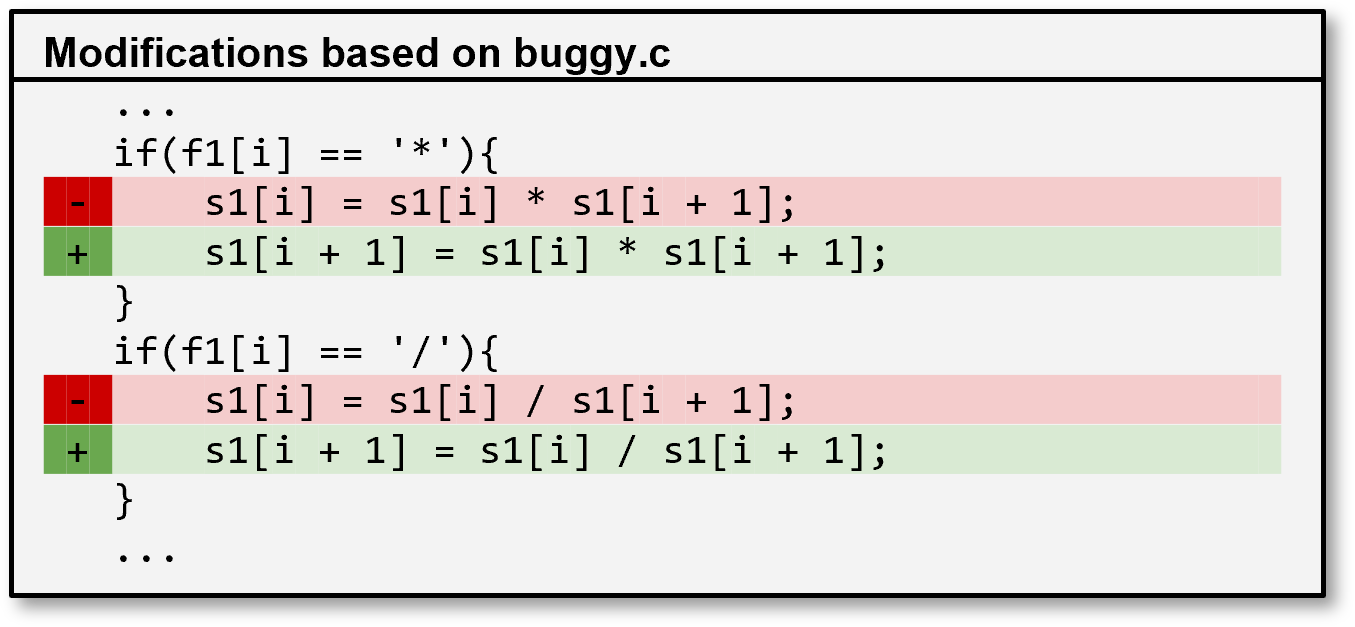}
     \end{subfigure}
    \caption{Bug fix example with the guidance of our peer solution strategy.}
    \label{fig:peer solution case}
    \vspace{-0.5cm}
\end{figure}

\subsubsection{Effectiveness of the selected similar peer solutions.}
We provide a specific example to demonstrate the effectiveness of our selection strategy using $\mathbf{PSM}$ score. As shown in Figure \ref{fig:peer solution case}, the original buggy code fails to increment the index of the assigned array element, while the selected peer solution correctly resolved this issue (the highlighted yellow lines), which has the highest $\textsc{match}_{tc}$ score (test cases pass match) and a $\textsc{match}_{df}$ score (data-flow match) higher than most other programs. The operation \textit{+1} offers a hint that helps the successful repair of this buggy code.
However, peer solutions selected by other strategies, such as BM25, do not include this similar change, and the model failed to fix this bug. This example affirms the superiority of our strategy in selecting semantically and syntactically related peer solutions.

\section{Threats to Validity}
\noindent \textbf{Internal.}
The internal threat to validity in this study is associated with the potential data leakage issue. This occurs when certain code snippets from students' correct submissions are included in the training data of ChatGPT or Code Llama. However, as students are expected to complete the assignments independently and the submissions in \ourdata{} are not uploaded to public repositories, we believe the risk of data leakage is minimal.

\noindent \textbf{External.}
The external threat comes from our evaluation dataset \ourdata{}. The results obtained in this dataset may not generalize to other programming assignment repair datasets. To alleviate this issue, we also evaluate \ourmodel{} on another generally used student assignment dataset ITSP to confirm the generalizability. Even though, our current findings are limited to assignments that can be solved by stand-alone programs, may not be applicable to assignments that involve the development of software systems or collaboration among multiple programmers.

\noindent \textbf{Construct.} The construct validity relates to the suitability of our evaluation metrics. We assess the quality of the generated programs by analyzing the proportion of test cases that successfully pass. Utilizing the tests as a proxy for correctness is common in the educational domain \cite{zhang2022repairing,Ahmed2021VerifixVR}. Our further analysis has confirmed that \ourmodel{} repairs the buggy code rather than generating an entirely new solution, supporting the feasibility of using pass rate-based metrics. However, it is important to note that validating program correctness through tests may not be as robust as formal verification methods.

\section{conclusion \& future work}
In this paper, we propose \ourmodel{}, an APR tool that leverages the LLM to repair semantic mistakes in advanced assignments with a novel peer solution selection strategy and a multi-source prompt generation method. We also build an advanced student assignment dataset named \ourdata{}, aiming to facilitate the evaluation of higher-level program assignment repair models. The evaluation results on both \ourdata{} and ITSP datasets prove that \ourmodel{} achieves the new state-of-the-art performance.
In the future, we will expand our dataset by incorporating assignments that involve software system development, like compiler systems, which demand a higher level of problem-solving and technical skills. Additionally, we will adapt our approach to providing effective feedback for tasks with varying difficulty levels and types. Ultimately, our goal is to enhance the quality of education in software development.

 

\bibliographystyle{ACM-Reference-Format}
\bibliography{sample-base}


\end{document}